\documentclass[sigconf]{acmart} %

\newcommand*\TechnicalReportURL{https://www.cl.cam.ac.uk/techreports/UCAM-CL-TR-1010.html}

\usepackage[caption=false,farskip=0pt,font=normalsize,labelfont=sf,textfont=sf]{subfig}

\usepackage{tabularx}
\newcolumntype{Y}{>{\raggedright\arraybackslash}X}

\usepackage{alltt}
\usepackage{url}
\usepackage{amsmath}
\usepackage{algorithm}
\usepackage{algpseudocodex}
\usepackage{colortbl}
\usepackage[inline,shortlabels]{enumitem}

\setlist{nosep,leftmargin=*}
\setlist[description]{style=unboxed,leftmargin=0pt}

\usepackage{tikz}
\usetikzlibrary{arrows,arrows.meta,backgrounds,calc,decorations.pathreplacing,patterns,positioning,shapes.symbols}
\usetikzlibrary{external}
\colorlet{rw}{yellow!25}
\colorlet{rx}{magenta!25}
\colorlet{reserved}{gray!25}
\pgfkeys{/pgf/arrow keys/scale=3}
\tikzset{
    legend/.style={
        every outer matrix/.append style={inner sep=0},
        nodes={anchor=west}
    },
    node distance=0,
    stack/.style={
        tape,
        tape bend height=.25em,
        minimum width=1.5cm,
        minimum height=#1
    },
    address space/.style={
        tape,
        tape bend height=.125em,
        minimum width=.5cm,
        minimum height=#1
    },
    addrspace/.style={
        tape,
        rotate=90,
        tape bend height=.125em,
        minimum width=2em,
        minimum height=#1
    },
    legend box/.style={
        draw,
        minimum width=1em,
        minimum height=1em,
        anchor=east
    }
}

\usepackage{hyperref}
\usepackage[capitalise]{cleveref}

\newenvironment{compactcontents}{\small}{}
\newenvironment{verbatimcontents}{%
  \footnotesize}{}

\copyrightyear{2026}
\acmYear{2026}
\setcopyright{cc}
\setcctype{by-nc-nd}
\acmConference[CCS '26]{Proceedings of the 2026 ACM SIGSAC Conference on Computer and Communications Security}{November 15--19, 2026}{The Hague, Netherlands}
\acmBooktitle{Proceedings of the 2026 ACM SIGSAC Conference on Computer and Communications Security (CCS '26), November 15--19, 2026, The Hague, Netherlands}
\acmDOI{10.1145/3830454.3846528}
\acmISBN{979-8-4007-2871-6/2026/11}

\begin{document}

\title{Efficient Linkage-Based Compartmentalization on CHERI}

\author{Dapeng Gao}
\affiliation{%
  \institution{University of Cambridge}
  \city{Cambridge}
  \country{UK}
}
\email{dapeng.gao@cl.cam.ac.uk}
\author{John Baldwin}
\affiliation{%
  \institution{Ararat River Consulting}
  \city{Ashland}
  \state{VA}
  \country{USA}
}
\email{john@araratriver.co}
\author{Jessica Clarke}
\affiliation{%
  \institution{University of Cambridge}
  \city{Cambridge}
  \country{UK}
}
\email{jessica.clarke@cl.cam.ac.uk}
\author{Nicholas C. Connolly}
\affiliation{%
  \institution{Arm Limited}
  \city{Cambridge}
  \country{UK}
}
\email{nick.connolly@arm.com}
\author{Brooks Davis}
\affiliation{%
  \institution{Capabilities Limited}
  \city{Llanwrda}
  \country{UK}
}
\additionalaffiliation{%
  \institution{University of Cambridge}
  \city{Cambridge}
  \country{UK}
}
\email{brooks@capabilitieslimited.co.uk}
\author{Franz A. Fuchs}
\affiliation{%
  \institution{University of Cambridge}
  \city{Cambridge}
  \country{UK}
}
\email{franz.fuchs@cl.cam.ac.uk}
\author{Alfredo Mazzinghi}
\affiliation{%
  \institution{University of Cambridge}
  \city{Cambridge}
  \country{UK}
}
\additionalaffiliation{%
  \institution{Capabilities Limited}
  \city{Llanwrda}
  \country{UK}
}
\email{alfredo.mazzinghi@cl.cam.ac.uk}
\author{Daniel Moghimi}
\affiliation{%
  \institution{Google}
  \city{Mountain View}
  \state{CA}
  \country{USA}
}
\email{danielmm@google.com}
\author{Peter Rugg}
\affiliation{%
  \institution{University of Cambridge}
  \city{Cambridge}
  \country{UK}
}
\authornotemark[2]
\email{peter.rugg@cl.cam.ac.uk}
\author{Domagoj Stolfa}
\affiliation{%
  \institution{University of Cambridge}
  \city{Cambridge}
  \country{UK}
}
\additionalaffiliation{%
  \institution{MSB Associates}
  \city{Belmont}
  \state{CA}
  \country{USA}
}
\email{domagoj.stolfa@cl.cam.ac.uk}
\author{Konrad Witaszczyk}
\affiliation{%
  \institution{University of Cambridge}
  \city{Cambridge}
  \country{UK}
}
\email{konrad.witaszczyk@cl.cam.ac.uk}
\author{Simon W. Moore}
\affiliation{%
  \institution{University of Cambridge}
  \city{Cambridge}
  \country{UK}
}
\email{simon.moore@cl.cam.ac.uk}
\author{Robert N. M. Watson}
\affiliation{%
  \institution{University of Cambridge}
  \city{Cambridge}
  \country{UK}
}
\authornotemark[2]
\email{robert.watson@cl.cam.ac.uk}
\renewcommand{\shortauthors}{Dapeng Gao et al.}

\begin{abstract}
We present an efficient \emph{linkage-based} model for in-process compartmentalization built on CHERI memory safety, which enables fine-grained compartmentalization of the entire UNIX user-space, scaling to 10K+ compartments on desktop systems.
The model's ``push-button'' compartmentalization along existing library boundaries regularly hosts 500+ compartments per process for large applications such as Chromium, far exceeding the number of concurrently available protection domains supported by other mechanisms (e.g., up to 16 for Intel MPK).
Custom policies can further subdivide libraries.
Of the thousands of C/C++ programs tested, only the V8 JavaScript engine required source-level adaptation ($<$300 lines of changed code concerning garbage collection and JIT compilation).

We implement the model for CHERI-extended versions of Armv8-A and RISC-V through support in the compiler toolchain and operating system.
Case studies illustrate the smooth delegation of memory between compartments, compartment-aware debugging and visualization, as well as extensibility to a complex managed language runtime, demonstrating the benefits of our single-address-space model.
We evaluate using multiple processors, including Arm's superscalar Morello and, notably, the first commercial CHERI-enabled RISC-V application core---Codasip's in-order dual-issue X730.

\end{abstract}




\maketitle

\tikzexternaldisable

\section{Introduction}

Compartmentalization solutions are fundamentally constrained by the available architectural primitives.
Techniques designed for commodity hardware often rely on features that were not originally intended for security, leading to programming models that are awkward to develop for, let alone to test and debug.
Although innovations in programming languages have lessened these difficulties, they introduce new friction points, such as the high cost of retrofitting existing code, performance degradation due to stylized code generation, and vastly expanded toolchain dependencies.

Expanding the design space of compartmentalization through architectural extensions, which has often been considered prohibitively expensive, now seems increasingly attractive to system vendors.
For example, Intel Memory Protection Keys (MPK) has enabled a family of efficient in-process compartmentalization systems by allowing user-space code to switch access permissions for groups of pages without kernel intervention, as discussed in \Cref{sec:related-work}.
However, many designs based on such extensions require static assignment (or costly dynamic reassignment) of protection to memory regions.
This makes them less suitable for \emph{data-dependent} sharing, which can require a combinatorial number of configurations: in our case study in \Cref{sec:tcpdump}, tcpdump routes packets among 100+ dissectors based on their contents, yielding $\gg 2^{100}$ possible access-control configurations.
To tackle this and other limitations, we set out to explore the alternative of building compartmentalization upon a \emph{capability system}---CHERI~\cite{watson_cheri_2012,woodruff_cheri_2014}.

CHERI extends pointers in existing ISAs to rich \emph{capabilities} that encode bounds and permissions in addition to integer memory addresses.
Since its inception more than a decade ago, OS and hardware support for it have evolved significantly.
On the CheriBSD operating system, CheriABI~\cite{davis_cheriabi_2019} elevates capabilities to first-class citizens in a UNIX environment, making it practical to run a substantial corpus of POSIX programs with fine-grained spatial memory safety.
CHERI heap temporal safety~\cite{xia_cherivoke_2019,filardo_cornucopia_2020,filardo_cornucopia_2024} has also matured sufficiently to be enabled by default on CheriBSD.
Arm developed the Morello prototype~\cite{grisenthwaite_arm_2023}, featuring a CHERI-enabled 2.5 GHz quad-core processor based on the Neoverse N1 that enables large-scale experiments.
Today, the CHERI ecosystem consists of a rich stack of memory-safe software, including a desktop environment, web servers, and more than 10K C/C++ packages~\cite{watson_cheri_2024}.

Building upon these advances, this work provides an efficient mechanism for developers to \emph{incrementally} enforce domain-specific compartmentalization policies.
It also serves as a foundation for future policy construction and auditing techniques, as illustrated in \Cref{fig:foundation} and discussed in \Cref{sec:future}.
In summary, we:
\begin{itemize}
    \item Present a model for compartmentalizing dynamically linked programs within a UNIX process, comprising:
    \begin{enumerate}[a)]
        \item Linkage-based policy representation designed for integration with existing compiler toolchains and build systems, and
        \item Efficient CHERI-based enforcement of security guarantees, integrated with existing OS components to preserve feature compatibility with minimal source-code disruption.
    \end{enumerate}
    \item Demonstrate that the model supports realistic software development workflows through compiler toolchain and debugger integration, and introduce CheriTree, a novel tool for tracing and visualizing run-time capability reachability.
    \item Implement the model for CHERI-extended versions of Armv8-A and RISC-V, and evaluate its performance across diverse applications, workloads, and processors.
\end{itemize}
In comparison to prior compartmentalization designs, including CHERI-based ones, our model contains the following key novelties:
\begin{itemize}
    \item Dynamic linking, underpinned by CHERI's capability-based memory safety, enables almost transparent in-address-space compartmentalization.
    \item This technique applies to complex constructs such as callbacks, C++ exceptions, JIT compilation, and garbage collection.
    \item The model scales efficiently to hundreds of compartments on multi-million-line applications with minimal source-code changes.
    \item The model integrates compartment-aware debugging, tracing, and visualization into established development workflows.
\end{itemize}

\tikzexternalenable
\begin{figure}[t]
\centering
\begin{compactcontents}
\begin{tikzpicture}
    \node[draw, minimum height=1.6em, minimum width=24em, outer sep=0] (frame0) {Scalable Security Policy Creation \& Auditing};

    \node[draw, below=of frame0.south west, anchor=north west, minimum height=1.6em, minimum width=11em, outer sep=0] (frame1-1) {Logic Hardening};
    \node[draw, below=of frame0.south east, anchor=north east, minimum height=1.6em, minimum width=13em, outer sep=0] (frame1-2) {Interface Hardening};

    \node[draw, below=of frame1-1.south west, anchor=north west, minimum height=1.6em, minimum width=24em, outer sep=0, fill=yellow!20] (frame2) {Linkage-Based Compartmentalization (This Work)};

    \node[draw, below=of frame2.south west, anchor=north west, minimum height=1.6em, minimum width=7em, outer sep=0] (frame3-1) {CheriABI~\cite{davis_cheriabi_2019}};
    \node[draw, below=of frame2.south east, anchor=north east, minimum height=1.6em, minimum width=17em, outer sep=0] (frame3-2) {Heap Temporal Safety~\cite{xia_cherivoke_2019,filardo_cornucopia_2020,filardo_cornucopia_2024}};

    \node[draw, below=of frame3-1.south west, anchor=north west, minimum height=1.6em, minimum width=24em, outer sep=0] (frame4) {Spatial Memory Safety~\cite{watson_cheri_2012,woodruff_cheri_2014}};
    \node[draw, below=of frame4, minimum height=1.6em, minimum width=24em, outer sep=0] (frame5) {Safe Speculative Execution};
\end{tikzpicture}
\end{compactcontents}
\caption{Envisioned stack of protections enabled by CHERI.}\label{fig:foundation}
\end{figure}
\tikzexternaldisable

\section{A CHERI Primer}\label{sec:cheri-primer}

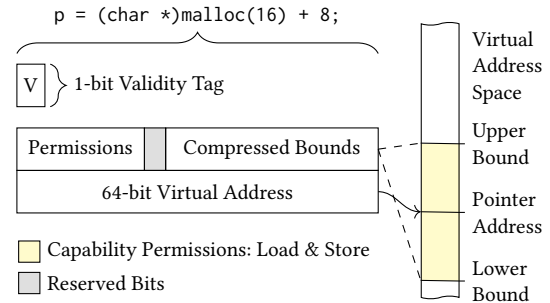
\begin{figure}[b]
\centering
\begin{compactcontents}
\begin{tikzpicture}
    \begin{scope}[local bounding box=lhs]
        \node[draw, minimum height=2em, anchor=north west, outer sep=0] (tag) {V};
        \draw[decorate, decoration={brace, amplitude=.75em}]
            ($(tag.north east)+(.25em,0)$) -- ($(tag.south east)+(0.25em,0)$)
            node[midway, right=.75em] {1-bit Validity Tag};
        \node[draw, minimum height=2em, minimum width=6em, below=1em of tag.south west, anchor=north west, outer sep=0] (perms) {Permissions\strut};
        \node[draw, minimum height=2em, minimum width=1em, right=of perms, outer sep=0, fill=reserved] (reserved) {};
        \node[draw, minimum height=2em, minimum width=10em, right=of reserved, outer sep=0] (bounds) {Compressed Bounds\strut};
        \node[draw, minimum height=2em, minimum width=17em, below=of perms.south west, anchor=north west, outer sep=0] (address) {64-bit Virtual Address};
        \matrix[legend, below=1em of address] (legend) {
            \node [draw, legend box, fill=rw] {}; & \node {Capability Permissions: Load \& Store}; \\
            \node [draw, legend box, fill=reserved] {}; & \node {Reserved Bits}; \\
        };
        \draw[decorate, decoration={brace, amplitude=1em}]
            ($(tag.north west)+(0,.5em)$) -- ($(tag.north -| bounds.east)+(0,0.5em)$)
            node[midway, above=1em] {\texttt{p = (char *)malloc(16) + 8;}};
    \end{scope}
    \begin{scope}
        \node [address space=13em, right=2em of lhs.south east, anchor=south west] (mem) {};
        \foreach \pos/\loc in {
            base/{15/16},
            addr/{11/16},
            top/{7/16}
        } {
            \coordinate (\pos) at ($(mem.north west)!\loc!(mem.south west)$);
            \coordinate (\pos bot) at ($(mem.north east)!\loc!(mem.south east)$);
        }
        \fill[rw] (base) rectangle (topbot);
        \foreach \pos/\label in {
            base/Lower\\Bound,
            addr/Pointer\\Address,
            top/Upper\\Bound
        } {
            \draw (\pos) -- (\pos bot) -- ++(.25em,0) node [right, align=left] {\label};
        }
        \node[draw, address space=13em, right=2em of lhs.south east, anchor=south west] {};
        \node[right=.25em of mem.north east, anchor=north west, align=left] {Virtual\\Address\\Space};
    \end{scope}
    \draw[->] (address.east) to [out=0, in=180] (addr);
    \draw[dashed]
        (bounds.east) -- (base)
        (bounds.east) -- (top);
\end{tikzpicture}
\end{compactcontents}
\caption{\texttt{malloc} returns a capability authorizing read and write permissions to a 16-byte allocation. Incrementing the capability by 8 changes the pointed-to address while retaining the original bounds and permissions.}\label{fig:capability}
\end{figure}

This section presents a simplified view of the CHERI ISA~\cite{watson_capability_2023}.
\Cref{fig:capability} illustrates how CHERI represents a C pointer, which can only be used to access a bounded region of memory with restricted privileges.
On a 64-bit implementation, a capability is represented as a 128-bit word in memory.
Its lower 64 bits contain the address, and the upper bits encode metadata including bounds and permissions.
The bounds are compressed like floating-point numbers, which leads to more rounding for wider bounds.
Some bits are reserved.
The register file is widened to hold capabilities, although it is only possible to perform arithmetic on the lower 64 bits of any register.

\tikzexternalenable
\tikzexternaldisable

Valid capabilities are distinguished from plain data down to the hardware level.
The hardware tracks whether each register and each 128-bit aligned word in memory contains a valid capability or not through a 1-bit tag associated with each location.
Each tag bit is always propagated together with its associated capability throughout the register file and memory hierarchy, providing atomicity.
New capabilities can be produced by special capability manipulation instructions, which guarantee the following invariant:
\begin{description}
    \item[Monotonicity] No sequence of instructions can derive a capability that is not a subset\footnote{A valid capability $c_1$ is a subset of capability $c_2$ when $c_2$ is valid, and the bounds and permissions of $c_1$ are subsets of those of $c_2$. For the sake of this definition, invalid capabilities are viewed as ``bottom'' elements that are trivially a subset of any capability.} of an existing capability.
\end{description}
This implies the following property:
\begin{description}
    \item[Unforgeability] No valid capability can be created ``out of thin air''. Instead, it must be derived from other valid capabilities.
\end{description}
Consequently, compared to the linear, integer-addressable virtual address space on conventional systems, CHERI offers an alternative perspective of memory as a \emph{graph} of bounded regions reachable from the register file via a chain of capability dereferences.
This approach is orthogonal to existing MMU-based, page-granular memory management but is much more dynamic and finer-grained.

\emph{Sealed entry} capabilities (also known as \emph{sentries}) are a special type of immutable and non-dereferenceable capability.
Only when installing it as the program counter via an indirect jump(-and-link) instruction can the processor \emph{unseal} a sentry into a mutable and dereferenceable form.
In CheriABI, function pointers are always sealed upon creation, and jump-and-link instructions always seal the return pointer.
Indirect control transfers through function pointers therefore begin at their sentry-designated entry points.
Sentries are also useful as opaque data pointers and unforgeable tokens.

\section{Abstract Design}

\tikzexternalenable
\begin{figure}[b]
\centering
\begin{compactcontents}
\begin{tikzpicture}
    \node[draw, minimum width=11em, align=center, fill=yellow!20] (source) {Source Files};

    \node[draw, below=1.6em of source, minimum width=11em] (frame0) {Compiler};
    \node[draw, below=1.6em of frame0, minimum width=11em] (frame1) {Static Linker};
    \node[draw, below=1.6em of frame1, minimum width=11em] (frame2) {Dynamic Linker};
    \node[draw, below=1.6em of frame2, minimum width=11em, align=center] (frame3) {CHERI and\\Compartmentalization\\TCB Protection};

    \node[draw, right=2em of frame0, minimum width=11em, fill=blue!20] (objects) {Relocatable Objects};
    \node[draw, right=2em of frame1, minimum width=11em, align=center, fill=yellow!20] (policy) {Compartmentalization\\Policy};

    \node[draw, right=2em of frame2, minimum width=11em, align=center, fill=blue!20] (exe) {Executables and\\Shared Libraries};

    \node[draw, right=2em of frame3, minimum width=11em, align=center] (space) {Instantiated\\Compartments};

    \draw[->] (source) to (frame0);

    \draw[->] (frame0) to (objects);
    \draw[->] (objects.south west) to (frame1.north east);
    \draw[->] (policy) to (frame1);

    \draw[->] (frame1.south east) to (exe.north west);
    \draw[->] (exe) to (frame2);

    \draw[->] (frame2.south east) to (space.north west);
    \draw[->] (space) to (frame3);

    \draw[decorate, decoration={brace, amplitude=.5em}]
        ($(frame1.south west)-(.25em,0)$) -- ($(source.north west)-(.25em,0)$)
        node[midway, xshift=-1em, rotate=90] {Build time};
    \draw[decorate, decoration={brace, amplitude=.5em}]
        ($(frame3.south west)-(.25em,0)$) -- ($(frame2.north west)-(.25em,0)$)
        node[midway, xshift=-1em, rotate=90] {Run time};

    \matrix[legend, above=.8em of objects] (legend) {
        \node [draw, legend box, fill=yellow!20] {}; & \node {Input Files}; \\
        \node [draw, legend box, fill=blue!20] {}; & \node {Build Artifacts}; \\
    };
\end{tikzpicture}
\end{compactcontents}
\caption{Pipeline of linkage-based compartmentalization.}\label{fig:flow}
\end{figure}
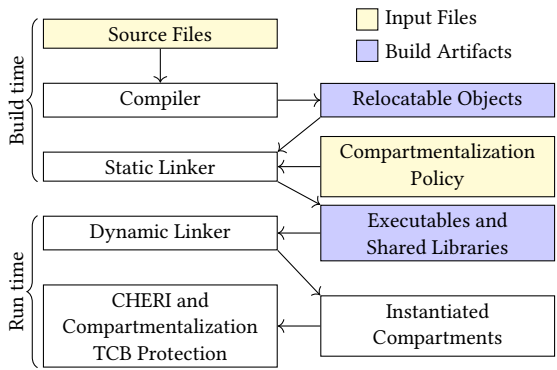
\tikzexternaldisable

\paragraph{Threat Model}\label{sec:threat-model}
Linkage-based compartmentalization applies to dynamically linked user-space programs in the UNIX environment, where the OS kernel, dynamic linker, C/C++ runtime, threading library, and a few highly privileged standard library functions form the Trusted Computing Base (TCB).\footnote{The TCB can itself be further hardened through compartmentalization: C++ exception handling already occurs separately from the C++ standard library, and we have begun isolating privileged \texttt{libc} components such as \texttt{malloc} and \texttt{stdio}.}
Attackers can exploit existing vulnerabilities in application code and insert backdoors through supply-chain attacks.
However, the compiler toolchain used to build all application code is trusted to produce ELF files that a) are well-formed and do not trigger parsing errors and b) faithfully apply the compartmentalization policy (see \Cref{sec:policy}), effectively excluding supply-chain attacks on build systems.
The hardware platforms are assumed to meet the CHERI ISA specification, whose soundness has been subject to formal verification~\cite{nienhuis_rigorous_2020,bauereiss_verified_2022}.
Micro-architectural side-channel attacks~\cite{lipp_meltdown_2018,kocher_spectre_2019} and physical attacks are out of scope.

\paragraph{Synopsis}
Our model operates in two phases as depicted by \Cref{fig:flow}.
During \emph{build time}, source files are compiled into relocatable objects, which are linked into executables or shared libraries.
Developers can supply the static linker with a \emph{compartmentalization policy} that dictates compartment composition and symbol accessibility in the output binaries (see \Cref{sec:policy}).
During \emph{run time}, the dynamic linker loads these binaries into memory and relocates them according to their policy-approved ELF section layouts and dynamic relocations.
Programs then execute under both standard CHERI memory safety protections and additional security guarantees provided by the compartmentalization TCB (see \Cref{sec:programming-model}).

\subsection{Compartmentalization Policy}\label{sec:policy}

We design our model to support effortless adoption and incremental refinement.
With no explicit policy, the executable and shared libraries form separate compartments, requiring essentially no developer effort.
With domain knowledge, developers can draw more effective security boundaries \emph{incrementally} by (a) using \emph{placement rules} to partition libraries into \emph{sub-libraries} and (b) using an \emph{access-control list} to authorize inter-compartment symbol references.

\tikzexternalenable
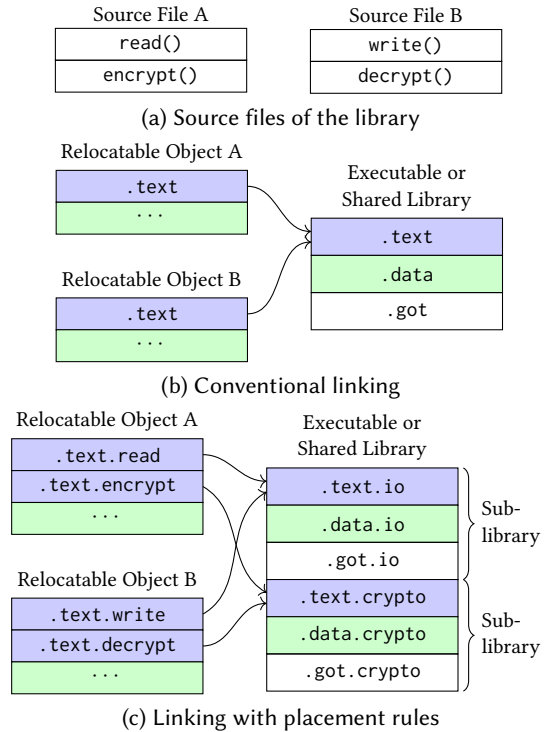
\begin{figure}[b]
\centering
\subfloat[Source files of the library]{
    \begin{compactcontents}
    \begin{tikzpicture}
    \begin{scope}[local bounding box=objA, name prefix=objA-]
        \node[draw, minimum height=1.5em, minimum width=9em, outer sep=0] (frame0) {\texttt{read()}};
        \node[draw, below=of frame0, minimum height=1.5em, minimum width=9em, outer sep=0] (frame1) {\texttt{encrypt()}};
        \node[above=of frame0] {Source File A};
    \end{scope}
    \begin{scope}[local bounding box=objB, name prefix=objB-]
        \node[draw, right=3em of objA, minimum height=1.5em, minimum width=9em, outer sep=0] (frame0) {\texttt{write()}};
        \node[draw, below=of frame0, minimum height=1.5em, minimum width=9em, outer sep=0] (frame1) {\texttt{decrypt()}};
        \node[above=of frame0] {Source File B};
    \end{scope}
\end{tikzpicture}
    \end{compactcontents}\label{fig:linking-sources}
}\\
\subfloat[Conventional linking]{
    \begin{compactcontents}
    \begin{tikzpicture}
    \begin{scope}[local bounding box=objA, name prefix=objA-]
        \node[draw, minimum height=1.5em, minimum width=9em, outer sep=0, fill=blue!20] (frame0) {\texttt{.text}};
        \node[draw, below=of frame0, minimum height=1.5em, minimum width=9em, outer sep=0, fill=green!20] (frame1) {$\cdots$};
        \node[above=of frame0] {Relocatable Object A};
    \end{scope}
    \begin{scope}[local bounding box=objB, name prefix=objB-]
        \node[draw, below=3em of objA, minimum height=1.5em, minimum width=9em, outer sep=0, fill=blue!20] (frame0) {\texttt{.text}};
        \node[draw, below=of frame0, minimum height=1.5em, minimum width=9em, outer sep=0, fill=green!20] (frame1) {$\cdots$};
        \node[above=of frame0] {Relocatable Object B};
    \end{scope}
    \begin{scope}[local bounding box=exe, name prefix=exe-]
        \node[draw, right=3em of objA-frame1, anchor=north west, minimum height=1.75em, minimum width=9em, outer sep=0, fill=blue!20] (frame0) {\texttt{.text}};
        \node[draw, below=of frame0, minimum height=1.75em, minimum width=9em, outer sep=0, fill=green!20] (frame1) {\texttt{.data}};
        \node[draw, below=of frame1, minimum height=1.75em, minimum width=9em, outer sep=0] (frame2) {\texttt{.got}};
        \node[above=of frame0, align=center] {Executable or\\Shared Library};

        \draw[->] (objA-frame0.east) to [out=0, in=180] ([yshift=.25em]frame0.west);
        \draw[->] (objB-frame0.east) to [out=0, in=180] ([yshift=-.25em]frame0.west);
    \end{scope}
\end{tikzpicture}
    \end{compactcontents}\label{fig:linking-conv}
}\\
\subfloat[Linking with placement rules]{
    \begin{compactcontents}
    \begin{tikzpicture}
    \begin{scope}[local bounding box=objA, name prefix=objA-]
        \node[draw, minimum height=1.5em, minimum width=9em, outer sep=0, fill=blue!20] (frame0) {\texttt{.text.read}};
        \node[draw, below=of frame0, minimum height=1.5em, minimum width=9em, outer sep=0, fill=blue!20] (frame1) {\texttt{.text.encrypt}};
        \node[draw, below=of frame1, minimum height=1.5em, minimum width=9em, outer sep=0, fill=green!20] (frame2) {$\cdots$};
        \node[above=of frame0] {Relocatable Object A};
    \end{scope}
    \begin{scope}[local bounding box=objB, name prefix=objB-]
        \node[draw, below=3em of objA, minimum height=1.5em, minimum width=9em, outer sep=0, fill=blue!20] (frame0) {\texttt{.text.write}};
        \node[draw, below=of frame0, minimum height=1.5em, minimum width=9em, outer sep=0, fill=blue!20] (frame1) {\texttt{.text.decrypt}};
        \node[draw, below=of frame1, minimum height=1.5em, minimum width=9em, outer sep=0, fill=green!20] (frame2) {$\cdots$};
        \node[above=of frame0] {Relocatable Object B};
    \end{scope}
    \begin{scope}[local bounding box=exe, name prefix=exe-]
        \node[draw, right=3em of objA-frame1, minimum height=1.75em, minimum width=9em, outer sep=0, fill=blue!20] (frame0) {\texttt{.text.io}};
        \node[draw, below=of frame0, minimum height=1.75em, minimum width=9em, outer sep=0, fill=green!20] (frame1) {\texttt{.data.io}};
        \node[draw, below=of frame1, minimum height=1.75em, minimum width=9em, outer sep=0] (frame2) {\texttt{.got.io}};
        \node[draw, below=of frame2, minimum height=1.75em, minimum width=9em, outer sep=0, fill=blue!20] (frame3) {\texttt{.text.crypto}};
        \node[draw, below=of frame3, minimum height=1.75em, minimum width=9em, outer sep=0, fill=green!20] (frame4) {\texttt{.data.crypto}};
        \node[draw, below=of frame4, minimum height=1.75em, minimum width=9em, outer sep=0] (frame5) {\texttt{.got.crypto}};
        \node[above=of frame0, align=center] {Executable or\\Shared Library};

        \draw[->] (objA-frame0.east) to [out=0, in=180] ([yshift=.25em]frame0.west);
        \draw[->] (objB-frame0.east) to [out=30, in=210] ([yshift=-.25em]frame0.west);

        \draw[->] (objA-frame1.east) to [out=-30, in=150] ([yshift=.25em]frame3.west);
        \draw[->] (objB-frame1.east) to [out=0, in=180] ([yshift=-.25em]frame3.west);

        \draw[decorate, decoration={brace, amplitude=.5em}]
            ($(frame0.north east)+(.25em,0)$) -- ($(frame2.south east)+(.25em,0)$)
            node[midway, xshift=2.25em, align=left] {Sub-\\library};
        \draw[decorate, decoration={brace, amplitude=.5em}]
            ($(frame3.north east)+(.25em,0)$) -- ($(frame5.south east)+(.25em,0)$)
            node[midway, xshift=2.25em, align=left] {Sub-\\library};
    \end{scope}
\end{tikzpicture}
    \end{compactcontents}\label{fig:linking-after}
}
\caption{Building a library with and without placement rules.}\label{fig:linking}
\end{figure}
\tikzexternaldisable

\paragraph{Placement Rules}
Shared libraries can sometimes be divided into smaller isolation units.
Suppose library A uses I/O only to log to one file.
Rather than granting its compartment full I/O privileges, developers can move the file-specific logging functions to compartment B and expose only them to A.
Compromising A then permits corruption of only the designated log file rather than arbitrary files.

Instead of manually splitting libraries, which would add major complications to build systems and package distribution, developers can refine the default compartments by supplying \emph{placement rules} to the compiler toolchain to emit a single shared library file that contains multiple \emph{sub-libraries}.
These rules describe a) the name of each sub-library and b) the symbols that belong to each sub-library.

\Cref{fig:linking} illustrates how to use placement rules to split a cryptographic library into multiple sub-libraries.
The library consists of two source files defining various functions, as depicted in \Cref{fig:linking-sources}.
The original build process involves compiling both source files into relocatable objects and linking them into a shared library, as illustrated in \Cref{fig:linking-conv}.
The following placement rules are applied to put the I/O and cryptographic functions into distinct sub-libraries:

\begin{verbatimcontents}
\begin{verbatim}
    {"compartments": {
      "crypto": {"symbols": ["encrypt", "decrypt"]},
      "io": {"symbols": ["read", "write"]}}}
\end{verbatim}
\end{verbatimcontents}
As shown in \Cref{fig:linking-after}, the functions are now regrouped into two sub-libraries in the output binary, but the overall structure of the build process remains unchanged.

\paragraph{Access-Control List (ACL)}
The other component of the policy is an ACL that defines which symbols each sub-library may reference and with what permissions.
Each ACL entry contains:
\begin{description}
    \item[Subject] A glob string identifying one or more sub-libraries that this ACL entry applies to.
    \item[Object] A set of glob strings identifying symbols that the subject is allowed to reference.
    \item[Permissions] A subset of $\{ \text{Read}, \text{Write}, \text{Execute} \}$ defining the rights that the subject has over the object.
\end{description}

\subsection{Programming Model}\label{sec:programming-model}

We integrate most linkage-based compartmentalization features into existing OS components and expose few new primitives.
In the general case, client code need not explicitly invoke any new APIs to be compartmentalized.
We also make design choices such that:
\begin{itemize}
    \item Programs built with a compartmentalization policy can still be run without compartmentalization.
    \item Programs built \emph{without} a policy can still be run \emph{with} compartmentalization using default per-library compartments.
    \item No compiler toolchain flag is required, apart from optionally specifying the policy, to build binaries for compartmentalization.
\end{itemize}
These strong compatibility properties yield the following benefits:
\begin{itemize}
    \item Existing build systems can be reused to produce binaries that work with or without compartmentalization.
    \item Diagnosing compartmentalization issues is simplified because the same binaries also run without compartmentalization.
    \item The same shared libraries can be used by processes with or without compartmentalization, minimizing system bloat.
\end{itemize}
To support programming models with more complex features such as just-in-time compilation and garbage collection, the compartmentalization TCB does expose a few highly privileged APIs (see \Cref{sec:setjmp}) for inspecting and manipulating its internal state.
\Cref{sec:v8} showcases their use by the V8 garbage collector.

\paragraph{Security Guarantees}
During compartment transitions, the following \emph{ABI hygiene} properties are enforced:
\begin{description}
    \item[Stack Safety] No compartment can access another compartment's stack frames except through explicitly shared capabilities.
    \item[Register Integrity] Callee-saved registers are restored to their pre-call values after a callee compartment returns.
    \item[Register Confidentiality] Registers not used for holding arguments or return values are cleared.
    \item[Control Flow Integrity] Control flow is ``well-bracketed'' so that it is impossible for a callee compartment to stash a return pointer and reuse it later to enter another compartment unexpectedly.
\end{description}
These properties ensure the integrity and confidentiality of inter-compartment control and data flow at the calling convention level, thereby establishing a limited form of \emph{mutual distrust}.

\paragraph{Trusted Compartments}\label{sec:trusted-compartments}
It is possible to opt out of ABI hygiene enforcement on certain compartment transitions for performance reasons.
By declaring a compartment as ``trusted'' in the policy, all transitions to it are elided.
Our implementation by default puts a number of stateless\footnote{Although statelessness is a desirable property for trusted compartments to have, it is neither a sufficient nor a necessary condition. The code must be audited to ensure that it cannot corrupt any caller's state.} standard library functions such as \texttt{memcpy} and \texttt{strcmp} into a trusted compartment.
We have also prototyped a feature that allows a set of compartments to be marked as ``mutually trusting'', effectively coalescing them into the same compartment and eliding transitions among them when possible.
Several benchmarks in \Cref{sec:eval} make use of this feature to improve performance.

\paragraph{POSIX Integration}
Our model aims to be \emph{orthogonal} to POSIX:
\begin{description}
    \item[Threading] Since compartments are associated with code alone, threads are not inherently associated with any one of them.
    \item[System Calls] Only system calls made through wrapper functions provided by the C runtime are accepted by the kernel. Thus, access to system calls is controlled in the same way as access to regular functions by the ACL (see \Cref{sec:syscalls}).
    \item[Signals] A signal triggers a compartment transition from the interrupted compartment to the signal handler's (see \Cref{sec:signal}).
\end{description}

\section{Implementation}\label{sec:mech}

We extend the CHERI LLVM toolchain to enforce compartmentalization policies (\Cref{sec:link}) and introduce a compartmentalization runtime of fewer than 2500 lines of code to CheriBSD to implement compartment transitions, language runtime support, and OS integration (\crefrange{sec:trampoline}{sec:signal}).
This section extensively uses terminology related to linking.
See the appendix for a primer.

\subsection{Compartmentalization Policy Enforcement}\label{sec:link}

\subsubsection{Applying Placement Rules}\label{sec:applying-placement-rules}

To keep the placement rules brief and aid the incremental partitioning of libraries into sub-libraries, not all functions need to be explicitly assigned to a sub-library by the placement rules.
When assembling the sub-libraries, the static linker first processes those explicitly assigned functions and then implicitly assigns the rest.
If the placement rules assign several functions from one relocatable object to different sub-libraries, then these functions must be in \emph{different} input sections.
However, the compiler places all functions in the same input section by default.
We therefore force each function to be in its own input section by always enabling the ``\texttt{-ffunction-sections}'' compiler flag, allowing the static linker to satisfy arbitrary placement rules.

\paragraph{Creating Sub-libraries}

The static linker first assigns the input sections that are explicitly mentioned by the placement rules to their respective sub-libraries.
It also generates per-sub-library synthetic output sections containing the GOT, PLT, etc.
Only input sections assigned to the same sub-library can be merged together, resulting in a set of output sections \emph{per sub-library}.
Each set is then assembled into a sub-library that contains its own code and data sections.
Some output sections such as symbol tables and non-PLT relocation tables are shared by all sub-libraries, which minimizes changes required in the dynamic linker and other tools such as debuggers.

\paragraph{Assigning Anonymous Objects}

Read-only data sections containing constants, such as anonymous strings or jump tables, must be placed in the same sub-library as the code that uses them because they are accessed through PC-relative addressing and hence must reside within the bounds of the PC (see \cref{sec:sub-libraries}).
The static linker therefore implicitly assigns them to the relevant sub-libraries, which is not only required for correctness but also avoids the need for placement rules to name these anonymous objects. 

\paragraph{Implicitly Assigning Sections}

The static linker then attempts to use a graph-based algorithm to assign the remaining sections.
Each static relocation creates an edge between the input section containing the instruction or data patched by the relocation and the input section containing the target symbol referenced by the relocation.
These edges partition the input sections into disjoint connected components.
If a connected component contains input sections already assigned to sub-libraries (whether explicitly or due to PC-relative accesses), and all such sections belong to the same sub-library, then its unassigned input sections are also assigned to that sub-library.
Finally, any remaining unassigned input sections are assigned to a default catch-all sub-library.

\subsubsection{ACL Enforcement}
For each static relocation, the static linker consults the ACL to determine if the requested reference is permitted.
If the reference is prohibited, a link-time error is emitted.
If the reference is permitted and the result of the relocation is a capability (e.g., a GOT entry), the static linker constrains the permissions of the resulting capability according to the ACL.
For example, references from sub-library A to a \emph{writable} data symbol defined in sub-library B will be authorized by a \emph{read-only} capability if the ACL \emph{only} permits read access to that data symbol from sub-library A.

\subsubsection{Isolating Sub-libraries}\label{sec:sub-libraries}
CheriABI already isolates shared libraries to some extent: each code pointer is a capability that is initialized at run time by the dynamic linker, which restricts the pointer's bounds to the memory image of the shared library to which the pointed-to code belongs.
This choice of bounds is sufficiently permissive to allow the use of direct branches for intra-library function calls but restrictive enough to prevent any code from using PC-relative addressing to call into another library.

Linkage-based compartmentalization brings the granularity of this isolation down to sub-libraries.
\emph{First}, the static linker inserts appropriate padding between sub-libraries to ensure that the rounded capability bounds for each sub-library do not overlap with other sub-libraries.
\emph{Second}, the dynamic linker now bounds all code pointers to the memory image of their containing sub-libraries rather than that of the entire shared library.
\emph{Third}, the static linker now emits PLT stubs and PLTGOT entries for inter-sub-library calls to non-preemptible functions in the same relocatable object, as opposed to the convention of using direct PC-relative branches.
This indirection is necessary due to the narrower bounds of the program counter making target functions in other sub-libraries no longer reachable via direct branches.
Note that these new PLTGOT entries do not use lazy binding and jump slot relocations like typical PLTGOT entries but are initialized using relative relocations.

\subsection{Trampoline}\label{sec:trampoline}

Recall from \Cref{sec:link} that a compartment can only call into another through function pointers initialized by the dynamic linker.
Each such function pointer is ``wrapped'' in a \emph{trampoline}:
when the dynamic linker creates a pointer to function $F$ when processing relocations or lazy binding requests, it generates a piece of TCB code $T_F$ known as a trampoline and returns a sentry pointing to it instead of $F$.\footnote{As an optimization, a function pointer need \emph{not} be wrapped in a trampoline if it is known to be a) never used for inter-compartment calls and b) never compared against other values by client code. This happens when, for example, a compartment binds to a preemptible function defined by itself that is not preempted at run time. Here, the resulting function pointer is stored in the PLTGOT, hence impossible to be assigned to a C variable, and it is only used for \emph{intra}-compartment calls.}
This sentry is returned every subsequent time $F$ is referenced so that all pointers to the same function compare equal.

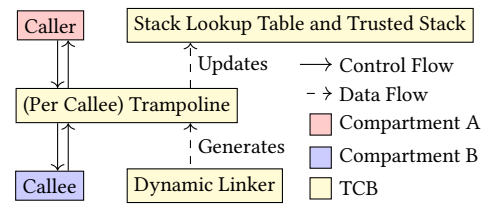
\begin{figure}[b]
\centering
\begin{compactcontents}
\begin{tikzpicture}
    \begin{scope}[local bounding box=picture]
        \matrix[row sep=2.125em, column sep={1em,between origins},
            column 1/.style={nodes={anchor=east}},
            column 2/.style={nodes={xshift=1em}},
            column 3/.style={nodes={anchor=west}}
        ] {
            \node[draw, fill=red!20] (caller) {Caller}; && \node[draw, fill=yellow!20] (trusted) {Stack Lookup Table and Trusted Stack}; \\
            & \node[draw, fill=yellow!20] (tramp) {(Per Callee) Trampoline}; & \\
            \node[draw, fill=blue!20] (callee) {Callee}; && \node[draw, fill=yellow!20] (linker) {Dynamic Linker}; \\
        };

        \draw[<-] ([xshift=-26]tramp.north) -- ([xshift=-26]tramp.north |- caller.south);
        \draw[->] ([xshift=-26]tramp.south) -- ([xshift=-26]tramp.south |- callee.north);

        \draw[->] ([xshift=-22]tramp.north) -- ([xshift=-22]tramp.north |- caller.south);
        \draw[<-] ([xshift=-22]tramp.south) -- ([xshift=-22]tramp.south |- callee.north);

        \draw[->, dashed] ([xshift=24]tramp.north) -- node[right] {Updates} ([xshift=24]tramp.north |- trusted.south);
        \draw[<-, dashed] ([xshift=24]tramp.south) -- node[right] {Generates} ([xshift=24]tramp.south |- linker.north);
    \end{scope}

    \matrix[legend, right=.75em of linker.south east, anchor=south west] {
        \draw[<-] (0,0) -- ++(-1.5em,0); & \node {Control Flow}; \\
        \draw[<-, dashed] (0,0) -- ++(-1.5em,0); & \node {Data Flow}; \\
        \node[legend box, fill=red!20] {}; & \node {Compartment A}; \\
        \node[legend box, fill=blue!20] {}; & \node {Compartment B}; \\
        \node[legend box, fill=yellow!20] {}; & \node {TCB}; \\
    }; 
\end{tikzpicture}
\end{compactcontents}
\caption{Trampolines mediate inter-compartment calls.}\label{fig:transition}
\end{figure}

Each trampoline has on the order of 100 instructions and is prefixed by a data header containing a pointer to the callee.
These instructions can be divided into a \emph{calling leg} and a \emph{returning leg}.
$T_F$ uses the same calling convention as $F$, and its calling leg forwards control to $F$ via a jump-and-link instruction, which points the link register to the returning leg through which $F$ can return to the caller.
As shown in \Cref{fig:transition}, the trampoline transparently transitions from compartment A to B and vice versa while updating the \emph{Trusted Stack} and \emph{Stack Lookup Table}, both crucial data structures for the compartmentalization TCB to be explained in more detail.

\tikzexternalenable
\tikzexternaldisable

By default, the calling leg of a trampoline includes a \emph{fast path} that elides transitions by directly tail-calling the callee if it belongs to the same compartment as the caller, which proves to be especially common in certain programs exemplified in \Cref{sec:grpc}.

\begin{figure}[t]
\centering
\begin{compactcontents}
\begin{tikzpicture}
    \begin{scope}[local bounding box=picture]
        \begin{scope}[local bounding box=regfile]
            \def\h{2.5}
            \def\w{1.3}
            \foreach \i/\label in {
                0/csp,
                1/pcc,
                2/c0,
                3/c1,
                4/c2,
                5/c3,
                6/c4,
                7/c5,
                8/c6,
                9/{...}
            } {
                \pgfmathsetmacro{\x}{\i * \w}
                \pgfmathsetmacro{\labelx}{\x + \w / 2}
                \pgfmathsetmacro{\labely}{\h / 2}
                \node[draw, rotate=90, minimum width=\h em, minimum height=\w em] (regfile\i) at (\labelx em, \labely em) {\texttt{\label}};
            }
        \end{scope}
        \node[left=of regfile, align=right] {Register File};

        \begin{scope}[local bounding box=caller]
            \node[draw, below=1.5em of regfile.south west, anchor=east,
                  addrspace=6em, tape bend top=none,
                  fill=rx] (pltgot) {};
            \node[right=of pltgot.south] (dots) {$\cdots$};
            \node[draw, right=of dots, anchor=north,
                  addrspace=10em, tape bend bottom=none,
                  fill=rx] (text) {};

            \coordinate (pc) at ($(text.north east)!1/2!(text.east)$);
            \coordinate (pcbot) at ($(text.north west)!1/2!(text.west)$);
            \draw (pc) -- (pcbot);

            \foreach \i in {1,...,5} {
                \coordinate (pltgot\i) at ($(pltgot.north east)!\i/6!(pltgot.south east)$);
                \coordinate (pltgotbot\i) at ($(pltgot.north west)!\i/6!(pltgot.south west)$);
                \draw (pltgot\i) -- (pltgotbot\i);
            }
            \coordinate (pltgotfocus) at ($(pltgotbot3)!1/2!(pltgotbot4)$);
        \end{scope}
        \node[left=of pltgot.north, align=right] {\texttt{.plt.got}\\Section};
        \node[right=of text.south, align=left] {\texttt{.text}\\Section};
        \draw[->] (regfile1.west) to [out=-30, in=150] (pc);
        \draw[dashed]
            (regfile1.west) -- (caller.north west)
            (regfile1.west) -- (caller.north east);

        \begin{scope}
            \node[below=2em of pltgot.south west, anchor=east, addrspace=16em] (tramp) {};
            \foreach \y in {0,1,2} {
                \foreach \pos/\offset in {base/1,entry/2,top/7} {
                    \pgfmathsetmacro{\loc}{(\y*7+\offset)/22}
                    \coordinate (tramp\y\pos) at ($(tramp.north east)!\loc!(tramp.south east)$);
                    \coordinate (tramp\y\pos bot) at ($(tramp.north west)!\loc!(tramp.south west)$);
                }
            }
            \fill[rx] (tramp1base) rectangle (tramp1topbot);
            \foreach \y in {0,1,2} {
                \foreach \pos/\offset in {base/1,entry/2,top/7}
                    \draw (tramp\y\pos) -- (tramp\y\pos bot);
                \fill[pattern=north east lines] (tramp\y base) rectangle (tramp\y entrybot);
                \fill[pattern=dots] (tramp\y entry) rectangle (tramp\y topbot);
            }
            \node[draw, below=2em of pltgot.south west, anchor=east, addrspace=16em] {};
            \coordinate (trampdatafocus) at ($(tramp1basebot)!1/2!(tramp1entrybot)$);
        \end{scope}
        \node[right=of tramp.south] {Trampoline Page};
        \draw[->] (pltgotfocus) to [out=-90, in=90] (tramp1entry);
        \draw[dashed]
            (pltgotfocus) -- (tramp1base)
            (pltgotfocus) -- (tramp1top);
    \end{scope}

    \node[draw, below=2em of tramp.north west, anchor=north west,
          minimum height=2em, minimum width=10em,
          fill=rx] (callee) {};
    \coordinate (target) at ($(callee.north west)!1/3!(callee.north east)$);
    \coordinate (targetbot) at ($(callee.south west)!1/3!(callee.south east)$);
    \draw (target) -- (targetbot);
    \draw[->] (trampdatafocus) to [out=-120, in=60] (target);
    \draw[dashed]
        (trampdatafocus) -- (callee.north west)
        (trampdatafocus) -- (callee.north east);
    \node[below=of callee] {Callee Code};
    \matrix[legend, below=.75em of picture.south east, anchor=north east] {
        \node[draw, legend box, fill=rx] {}; & \node[align=left] {Capability Permissions:\\Read \& Execute}; \\
        \node[draw, legend box, pattern=north east lines] {}; & \node {Trampoline: Data Header}; \\
        \node[draw, legend box, pattern=dots] {}; & \node {Trampoline: Instructions}; \\
    };
\end{tikzpicture}
\end{compactcontents}
\caption{The capability graph surrounding a trampoline.}\label{fig:trampoline}
\end{figure}
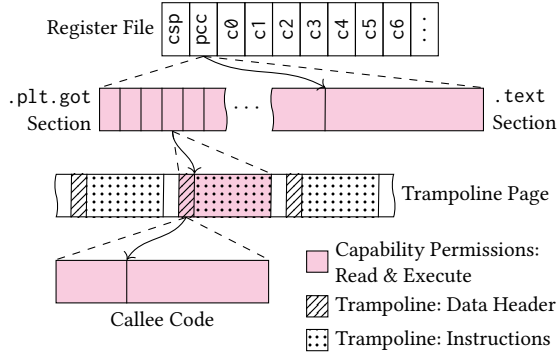

As symbol resolution proceeds, trampolines are laid out contiguously, with suitable padding, on writable and executable pages mapped by the dynamic linker.
While the use of such pages is normally frowned upon, there is little risk in doing so when leveraging CHERI's fine-grained memory protection in conjunction with existing Page Table Entry (PTE) permissions.
\Cref{fig:trampoline} shows a Morello register file whose Program Counter Capability\footnote{The program counter (PC) has been extended to a CHERI capability with bounds etc.} (PCC) covers both the PLTGOT and \texttt{.text}, as is the case for typical programs.
The PCC grants Read and Execute permissions, but because the PLTGOT is mapped onto pages that lack the execute PTE permission, the MMU would still stop any attempt to execute it.

\tikzexternalenable
\tikzexternaldisable

The figure then shows a PLTGOT entry pointing not to the callee but to a trampoline.
This capability only has Read and Execute permissions and hence cannot be used to modify the trampoline despite it being mapped onto a writable page.
Note that unlike the previous case, no \emph{architectural} feature prevents the data header of a trampoline from being executed.
Instead, the trampoline is carefully designed such that under no circumstances should this happen as long as only sentries pointing to its entry point as well as return pointers to its returning leg are exposed to client code.

\paragraph{Compiler Toolchain Changes}
When taking a reference to a non-preemptible function in the same compartment, the compiler may emit code to derive a function pointer directly from the PC.
Such a function pointer would \emph{not} be wrapped in a trampoline and would hence bypass compartment transitions if called from another compartment.
To prevent this, the compiler is modified to never derive function pointers from the PC but always load them from the GOT.

\subsection{Stack-Frame Isolation}\label{sec:stack-safety}

To prevent any compartment from:
\begin{itemize}
    \item Reading from the stack frames of other compartments, including deallocated frames that may have stale pointers, or
    \item Writing to the stack frames of other compartments
\end{itemize}
without explicit authorization by the programmer, our model assigns a dedicated stack to each compartment on each thread and switches the execution stack during compartment transitions.
As usual, this should all be transparent to client code, which may continue to assume the stack model of the C abstract machine.
But this implies that a \emph{contiguous abstract stack} can be reconstructed from the split stacks.
For this purpose, the compartmentalization TCB maintains two data structures per thread whose lifetimes are managed in conjunction with the threading library:
\begin{description}
    \item[Stack Lookup Table] For each compartment, record the top of its stack during the latest transition into or out of it.
    \item[Trusted Stack] Track a sequence of \emph{trusted frames} corresponding to the stack of inter-compartment calls so far. Each frame records the top of the calling compartment's stack during its \emph{previous} transition.
\end{description}

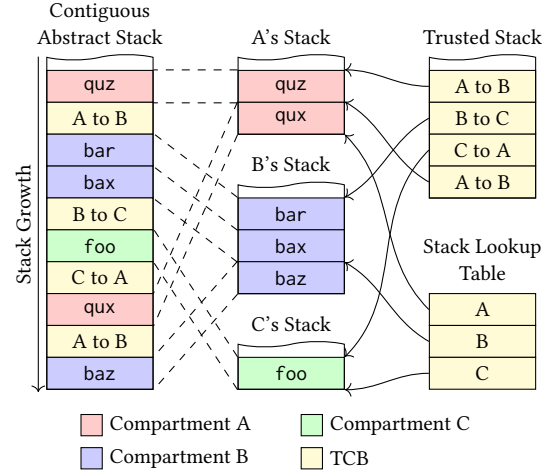
\begin{figure}[t]
\centering
\begin{compactcontents}
\begin{tikzpicture}
    \def\h{1.5}
    \def\w{5}
    \begin{scope}[local bounding box=stacks]
    \begin{scope}[name prefix=stack-]
        \foreach \i/\label/\background in {
            9/\texttt{quz}/red!20,
            8/A to B/yellow!20,
            7/\texttt{bar}/blue!20,
            6/\texttt{bax}/blue!20,
            5/B to C/yellow!20,
            4/\texttt{foo}/green!20,
            3/C to A/yellow!20,
            2/\texttt{qux}/red!20,
            1/A to B/yellow!20,
            0/\texttt{baz}/blue!20
        } {
            \pgfmathsetmacro{\y}{\i * \h}
            \pgfmathsetmacro{\labelx}{0}
            \pgfmathsetmacro{\labely}{\y}
            \node[draw, fill=\background, minimum width=\w em, minimum height=\h em] (frame\i) at (\labelx em, \labely em) {\label};
        }
        \node[draw, below=.125em of frame0.south, anchor=south, stack=15.75em, minimum width=5em, tape bend bottom=none] (stack) {};
        \draw[->] ([xshift=-.375em]stack.north west) -- ([xshift=-.375em]stack.south west) node[midway, above, rotate=90]{Stack Growth};
        \node[above=.25em of stack.north, anchor=south, align=center] {Contiguous\\Abstract Stack};
    \end{scope}
    \begin{scope}[name prefix=stackA-]
        \foreach \i/\label/\background in {
            1/\texttt{quz}/red!20,
            0/\texttt{qux}/red!20
        } {
            \pgfmathsetmacro{\y}{\i * \h}
            \pgfmathsetmacro{\labelx}{9}
            \pgfmathsetmacro{\labely}{\y + 12}
            \node[draw, fill=\background, minimum width=\w em, minimum height=\h em] (frame\i) at (\labelx em, \labely em) {\label};
        }
        \node[draw, below=.125em of frame0.south, anchor=south, stack=3.75em, minimum width=5em, tape bend bottom=none] (stack) {};
        \node[above=.25em of stack.north, anchor=south] {A's Stack};
        \draw[dashed] (frame1.north west) -- (stack-frame9.north east);
        \draw[dashed] (frame1.south west) -- (stack-frame9.south east);
        \draw[dashed] (frame0.north west) -- (stack-frame2.north east);
        \draw[dashed] (frame0.south west) -- (stack-frame2.south east);
    \end{scope}
    \begin{scope}[name prefix=stackB-]
        \foreach \i/\label/\background in {
            2/\texttt{bar}/blue!20,
            1/\texttt{bax}/blue!20,
            0/\texttt{baz}/blue!20
        } {
            \pgfmathsetmacro{\y}{\i * \h}
            \pgfmathsetmacro{\labelx}{9}
            \pgfmathsetmacro{\labely}{\y + 4.5}
            \node[draw, fill=\background, minimum width=\w em, minimum height=\h em] (frame\i) at (\labelx em, \labely em) {\label};
        }
        \node[draw, below=.125em of frame0.south, anchor=south, stack=5.25em, minimum width=5em, tape bend bottom=none] (stack) {};
        \node[above=.25em of stack.north, anchor=south] {B's Stack};
        \draw[dashed] (frame2.north west) -- (stack-frame7.north east);
        \draw[dashed] (frame2.south west) -- (stack-frame7.south east);
        \draw[dashed] (frame1.south west) -- (stack-frame6.south east);
        \draw[dashed] (frame0.north west) -- (stack-frame0.north east);
        \draw[dashed] (frame0.south west) -- (stack-frame0.south east);
    \end{scope}
    \begin{scope}[name prefix=stackC-]
        \foreach \i/\label/\background in {
            0/\texttt{foo}/green!20
        } {
            \pgfmathsetmacro{\y}{\i * \h}
            \pgfmathsetmacro{\labelx}{9}
            \pgfmathsetmacro{\labely}{\y}
            \node[draw, fill=\background, minimum width=\w em, minimum height=\h em] (frame\i) at (\labelx em, \labely em) {\label};
        }
        \node[draw, below=.125em of frame0.south, anchor=south, stack=2.25em, minimum width=5em, tape bend bottom=none] (stack) {};
        \node[above=.25em of stack.north, anchor=south] {C's Stack};
        \draw[dashed] (frame0.north west) -- (stack-frame4.north east);
        \draw[dashed] (frame0.south west) -- (stack-frame4.south east);
    \end{scope}
    \begin{scope}[name prefix=stackT-]
        \foreach \i/\label/\background in {
            3/A to B/yellow!20,
            2/B to C/yellow!20,
            1/C to A/yellow!20,
            0/A to B/yellow!20
        } {
            \pgfmathsetmacro{\y}{\i * \h}
            \pgfmathsetmacro{\labelx}{18}
            \pgfmathsetmacro{\labely}{\y + 9}
            \node[draw, fill=\background, minimum width=\w em, minimum height=\h em] (frame\i) at (\labelx em, \labely em) {\label};
        }
        \node[draw, below=.125em of frame0.south, anchor=south, stack=6.75em, minimum width=5em, tape bend bottom=none] (stack) {};
        \node[above=.25em of stack.north, anchor=south] {Trusted Stack};
        \draw[->] (frame3.west) to [out=180, in=0] (stackA-frame1.north east);
        \draw[->] (frame2.west) to [out=210, in=30] (stackB-frame2.north east);
        \draw[->] (frame1.west) to [out=225, in=45] (stackC-frame0.north east);
        \draw[->] (frame0.west) to [out=150, in=-30] (stackA-frame0.north east);
    \end{scope}
    \begin{scope}[name prefix=lookup-]
        \foreach \i/\label/\background in {
            2/A/yellow!20,
            1/B/yellow!20,
            0/C/yellow!20
        } {
            \pgfmathsetmacro{\y}{\i * \h}
            \pgfmathsetmacro{\labelx}{18}
            \pgfmathsetmacro{\labely}{\y}
            \node[draw, fill=\background, minimum width=\w em, minimum height=\h em] (frame\i) at (\labelx em, \labely em) {\label};
        }
        \node[above=.25em of frame2.north, anchor=south, align=center] {Stack Lookup\\Table};
        \draw[->] (frame2.west) to [out=135, in=-45] (stackA-frame0.south east);
        \draw[->] (frame1.west) to [out=150, in=-30] (stackB-frame1.south east);
        \draw[->] (frame0.west) to [out=180, in=0] (stackC-frame0.south east);
    \end{scope}
    \end{scope}
    \matrix[legend, below=.75em of stacks] {
        \node[legend box, fill=red!20] {}; & \node {Compartment A}; &[2em] \node[legend box, fill=green!20] {}; & \node {Compartment C}; \\
        \node[legend box, fill=blue!20] {}; & \node {Compartment B}; & \node[legend box, fill=yellow!20] {}; & \node {TCB}; \\
    }; 
\end{tikzpicture}
\end{compactcontents}
\caption{Function \texttt{quz} initiates a chain of inter- and intra-compartment calls through \texttt{bar}, \texttt{bax}, \texttt{foo}, \texttt{qux}, and \texttt{baz}, causing the compartmentalization TCB to update the Trusted Stack and Stack Lookup Table.}
\label{fig:stack}
\end{figure}

During an inter-compartment call from compartment X to Y, the trampoline updates both data structures as follows:
\begin{enumerate}
    \item Push a new trusted frame for the ``X to Y'' transition.
    \item Copy X's entry in the Stack Lookup Table to that trusted frame.
    \item Update the entry to the current top of X's stack.
\end{enumerate}
The trampoline can then install the value in Y's entry in the Stack Lookup Table as the execution stack before completing the transition.
Later, upon returning from Y to X, the returning leg of the trampoline reverses the steps above, restoring the execution stack back to X's in the process.
As an example, the right side of \Cref{fig:stack} shows the configuration of several concrete stacks after a chain of calls.
This can be mapped to the contiguous abstract stack on the left, which consists of call frames interspersed with TCB-managed trusted frames at compartment boundaries.

\tikzexternalenable
\tikzexternaldisable

One implementation detail is worth highlighting: Morello and RISC-V use different architectural features to restrict access to the Trusted Stack and Stack Lookup Table to the compartmentalization TCB.
The Morello implementation uses the ``Executive'' and ``Restricted'' modes that its ISA introduces:
trampolines are run in Executive mode, granting access to additional stack and thread pointer registers, while compartment code is run in Restricted mode, forbidding access to Executive mode registers.
RISC-V adds a new ``UTIDC'' control and status (CSR) register, readable by user-space code---both the compartmentalization TCB and compartments---but writable only by the kernel.
By writing a sealed capability to this register on process and thread creation, and granting only the trampolines and compartmentalization TCB the ability to unseal this capability, the same protection can be provided.

\paragraph{ABI Changes}
In the previous version of CheriABI, fixed arguments that do not fit into the argument registers are spilled onto the stack, and the callee reaches into this part of the \emph{caller's} frame to access these arguments.
This calling convention no longer works when stack safety is enforced.
CheriABI is therefore modified such that the caller explicitly passes a bounded capability that covers the arguments which have been spilled into memory, whether fixed or variadic, and the callee can access these arguments relative to this capability rather than the stack pointer.

\subsection{Register Integrity and Confidentiality}\label{sec:reg-confidentiality}

\paragraph{Integrity}
To prevent callee-saved registers from being corrupted by a callee compartment, the calling leg of each trampoline saves all of them in the trusted frame that it pushes, while the returning leg of the trampoline pops the trusted frame and restores them.

\paragraph{Confidentiality}
The compiler and static linker together emit a new ELF section that encodes the \emph{signature} of each global function in terms of which registers are used for arguments and return values, respectively.
When the dynamic linker generates the trampoline of any function, it refers to that function's signature and emits instructions to clear exactly those non-argument registers in the calling leg and non-return value registers in the returning leg.

It is important that a signature is emitted for each function referenced by the dynamic symbol table regardless of whether it is defined in that shared library so that any disagreement between the caller and callee on the signature of a function can be detected.
Consider the following example where compartment A defines function \texttt{foo} and compartment B calls it with a different signature:
\begin{description}
    \item[Compartment A] Defines \texttt{int foo(int, void *);}
    \item[Compartment B] Declares \texttt{int foo(int);}
\end{description}
If only compartment A's signature were taken into account, then \texttt{foo}'s trampoline would leave two argument registers uncleared.
The caller in compartment B, believing \texttt{foo} to have only one argument, would then unknowingly leak the content of one register.

The signature adopted by the trampoline must, therefore, depend on both parties.
The naive solution is to require that all callers and the callee agree on the number of argument and return value registers.
But this is incompatible with existing practice.
An example is the multiple function types permitted for POSIX signal handlers:
\begin{enumerate}
    \item \texttt{void (*)(int)}
    \item \texttt{void (*)(int, siginfo\_t *, void *)}
\end{enumerate}
Some OSes always invoke signal handlers as if they are of type 2, despite this being undefined behavior, since they assume this to work under any reasonable ABI.
To support this type of usage, callers are allowed to declare more argument registers than does the callee, but they must not declare more return value registers.

\subsection{Well-Bracketed Control Flow}\label{sec:cfi}

The returning legs of all trampolines are interchangeable: they contain identical code and rely solely on protected data in the Stack Lookup Table and Trusted Stack to restore the caller's state.
This design therefore enforces well-bracketed inter-compartment control flow regardless of which particular returning leg is entered.

\subsection{Tail-Calls}\label{sec:tail-calls}

Suppose function \texttt{foo} in compartment X calls into compartment Y, which in turn \emph{tail-calls} function \texttt{bar} in compartment Z.
A naive implementation would result in two back-to-back trusted frames, ``X to Y'' and ``Y to Z'', between \texttt{foo}'s and \texttt{bar}'s call frames on the contiguous abstract stack.
But both trusted frames should ideally be collapsed into one ``X to Z'' frame so that it appears as if \texttt{foo} called \texttt{bar} directly.
Otherwise, a program that continuously makes inter-compartment tail-calls would overflow the Trusted Stack with extraneous frames, even though it might never cause a stack overflow when run without compartmentalization.

The calling leg of each trampoline reliably detects inter-compartment tail-calls: when the return address points to the returning leg of the trampoline that pushed the topmost trusted frame.
If this is the case, no new trusted frame is pushed so that stack space usage stays constant.
Instead, the topmost trusted frame is modified in place to reflect the new identity of the callee compartment.

\subsection{\texttt{setjmp}/\texttt{longjmp} and C++ Exceptions}\label{sec:setjmp}

Despite breaking the well-bracketedness of control flow, \texttt{setjmp} and \texttt{longjmp} must be supported by any realistic compartmentalization mechanism due to their wide presence in extant code.
In our model, the definition of \texttt{longjmp} naturally extends to forcefully unwinding the contiguous abstract stack, destroying all intermediate call frames and trusted frames.
To facilitate this, the compartmentalization TCB exposes two privileged APIs\footnote{The names of the APIs mentioned in this section are stylized for better readability and are not the names used in the actual implementation.}:
\begin{description}
    \item[GetTopTrustedFrame()] Return a capability to the topmost trusted frame on the Trusted Stack.
    \item[UnwindTrustedStack(\texttt{tf})] Iteratively pop trusted frames until \texttt{tf} is reached, each time updating the Stack Lookup Table similarly to what the returning leg of a trampoline would do.
\end{description}
Initially, \texttt{setjmp} calls \texttt{GetTopTrustedFrame} and stores the return value in the jump buffer.
Later, \texttt{longjmp} calls \texttt{UnwindTrustedStack} with this value to reverse through a series of compartment transitions until reaching the trusted frame that initially called \texttt{setjmp}.

The C++ runtime implements exceptions by walking through each call frame from where the exception is thrown to where it is caught.
The compartmentalization TCB offers functional support via the following privileged APIs:
\begin{description}
    \item[IsTrampoline(\texttt{pc}, \texttt{tf})] Return a Boolean indicating whether code pointer \texttt{pc} points to the returning leg of the trampoline that pushed trusted frame \texttt{tf}.
    \item[GetNextTrustedFrame(\texttt{tf})] Return the address and content of the parent trusted frame of \texttt{tf}.
\end{description}
\Cref{alg:libunwind} illustrates this process with lines 4 and 5 representing newly added code.\footnote{Existing functions in the C++ runtime are italicized.}
Upon receiving an exception, the C++ runtime steps through the call frames, repeatedly calling \texttt{IsTrampoline} to determine whether the next call frame belongs to another compartment and is hence interposed by a trampoline.
When this occurs, it calls \texttt{GetNextTrustedFrame} to obtain the address and machine state of that call frame.
Finally, the C++ runtime unwinds the contiguous abstract stack by calling \texttt{UnwindTrustedStack} and resumes execution by jumping to the exception handler.
\begin{algorithm}
\caption{Stack Walking for C++ Exception Handling}\label{alg:libunwind}
\begin{compactcontents}
\begin{algorithmic}[1]
\State $s \gets \operatorname{\textit{GetMachineState}}()$
\Repeat
    \State $(s, end?) \gets \operatorname{\textit{StepCallFrame}}(s)$
    \If{$\lnot end? \land \operatorname{IsTrampoline}(s.\mathrm{pc}, s.\mathrm{trusted\_frame})$}
        \State $s \gets \operatorname{GetNextTrustedFrame}(s.\mathrm{trusted\_frame})$
    \EndIf
\Until{$end?$}
\end{algorithmic}
\end{compactcontents}
\end{algorithm}

\paragraph{ABI Changes}
The pointer to an exception handler should not be wrapped in a trampoline as it is a resumption point of a function.
In previous versions of CheriABI, the dynamic linker could not distinguish such pointers from function pointers, and so could not omit the trampoline when creating them.
This is solved by introducing distinct static relocations for ``code pointers''
(versus existing relocations for normal function pointers)
emitted in the relocatable object files by the compiler and assembler.
The distinction between relocations that should and should not result in trampolines
is then propagated to the dynamic linker via differing dynamic relocations.
This same mechanism is used to support GNU C's ``computed goto''.

\subsection{System Calls}\label{sec:syscalls}

The vast majority of programs issue system calls by calling wrappers implemented by the C runtime, which in turn use a trap instruction to call the kernel.
Our model makes this the \emph{only} legal way for compartmentalized code to issue system calls: attempting to use the trap instruction outside a wrapper causes an error.
To achieve this, the compartmentalization TCB tags all function pointers to these wrappers with an unforgeable \emph{user-defined} permission bit.
The kernel, upon receiving a system call, first checks this bit on the user-space PCC and proceeds only if it is set.
The wrapper symbols remain subject to the ACL enforcement described in \Cref{sec:link}.

\subsection{Signals}\label{sec:signal}

The kernel delivers a signal by calling a function pointer to the signal handler.
Since function pointers are wrapped in trampolines, they would normally cause a compartment transition from the interrupted compartment to the signal handler's.
However, this would not work if an interrupt occurs during a critical section of the compartmentalization TCB when the Trusted Stack and Stack Lookup Table are temporarily inconsistent with the machine state (e.g., in the middle of a trampoline).
Therefore, the compartmentalization TCB registers a centralized handler for all signals and interposes all calls to \texttt{sigaction}, similarly to what \texttt{pthread} does.
This centralized handler is responsible for completing the work of any critical sections before dispatching the signal to the actual handler.
To make this possible, each critical section is carefully designed for reentrancy or, as a last resort, blocks all signals.

\section{Limitations}\label{sec:limitations}

System interfaces with powerful ambient authorities impose inherent limitations:
\texttt{setjmp} exposes the machine state of the calling compartment;
signal handlers receive access to the machine context of interrupted code;
integer file descriptors are process-granular;
several APIs like \texttt{dlopen}, \texttt{ptrace}, \texttt{dl\_iterate\_phdr}, and \texttt{elf\_aux\_info} grant wide-reaching authorities to the process.

We expect the following engineering limitations to be resolved.
Assembly-generated ``wild'' function pointers do not cause compartment transitions and function pointer callers currently do not check for this.
Toolchain-emitted function signatures can be incorrect in the presence of undefined behavior (even if unreachable).
Thread-local storage is not compartmentalized yet, which would be achieved by requiring the thread control block to expose \emph{not} the direct backing storage of all thread-local variables, but a GOT-like structure where each entry contains a bounded pointer to an object in the backing storage.
When compartmentalization is enabled, the dynamic linker would create a dedicated thread control block for each compartment and only populate those entries for thread-local variables that the compartment is permitted to access.

\section{Security Analysis}\label{sec:security-analysis}

The following properties are provided by baseline CheriABI:

\begin{description}
    \item[Spatial Memory Safety]
    Authority to memory objects, including global data, stack objects, executable mappings, and heap allocations, is conveyed through capabilities rather than a global mapping from compartments to address ranges (see \Cref{sec:cheri-primer}).
    Attackers cannot gain access to objects merely by guessing their addresses.

    \item[Temporal Memory Safety]
    CheriBSD's temporal-safety mechanisms revoke outstanding capabilities to regions that are freed or unmapped before their storage or virtual addresses are reused~\cite{xia_cherivoke_2019,filardo_cornucopia_2020,filardo_cornucopia_2024}.
    This prevents a stale capability from regaining authority when a heap allocation is recycled or a virtual-address range is remapped, even when the capability has been delegated across compartment boundaries.
    The same guarantee applies to code and global data: unloading a shared library or retiring a JIT-code mapping revokes capabilities into that region, preventing stale pointers from accessing or executing a later mapping at the same address.

    \item[Control-Flow Safety]
    The dynamic linker bounds each code pointer to its target compartment's memory image (see \Cref{sec:sub-libraries}).
    CHERI monotonicity prevents compromised code from widening these bounds to cover another compartment.
    Function and return pointers are sentries that can transfer control only to their designated addresses (see \Cref{sec:cheri-primer}).
    A compromised compartment may therefore reuse jump-oriented programming gadgets within itself, but it cannot extend a gadget chain into another compartment by retargeting a sentry to enter the middle of a function.
\end{description}

The present work additionally enforces the following properties:

\begin{description}
    \item[Compartmentalization Policy Enforcement]
    The static linker rejects prohibited references and records ACL-derived permission constraints for permitted ones to code and global data (see \Cref{sec:policy}), while the dynamic linker applies these constraints when constructing capabilities and bounds code capabilities to their containing sub-libraries (see \Cref{sec:link}).
    Capabilities to stack, heap, and JIT-compiled objects are constrained dynamically due to CheriABI.
    This permits data-dependent sharing without knowing every eventual recipient when the allocation is created, as illustrated by tcpdump's recursive packet decoding in \Cref{sec:tcpdump}.

    \item[Mediation of Compartment Transitions]
    Inter-compartment function references are obtained through the GOT and wrapped in trampolines so that calls cannot bypass compartment transitions (see \Cref{sec:trampoline}).
    The kernel accepts system calls only from approved C-runtime wrappers carrying an unforgeable permission, whose symbols remain governed by the ACL (see \Cref{sec:syscalls}).
    Signals are first delivered to a centralized handler that completes any interrupted TCB critical section before transitioning to the application handler (see \Cref{sec:signal}).
    Together, these mechanisms prevent attacks based on bypassing a transition with a function pointer, issuing a raw system-call trap to evade policy, or interrupting a transition while its protected state is temporarily inconsistent.

    \item[Stack Safety]
    Each thread uses a distinct execution stack for every compartment, and trampolines switch stacks during compartment transitions (see \Cref{sec:stack-safety}).
    Rather than receiving a capability spanning its caller's frames, a callee whose arguments reside in memory receives a separate capability bounded to only the argument area.
    This prevents a compromised callee from inspecting or corrupting unrelated caller frames while preserving explicit sharing of stack objects required by the C programming language.

    \item[Register Integrity and Confidentiality]
    Trampolines save callee-saved registers in protected trusted frames and restore them on return, preserving the caller's register state across a mutually distrusting call.
    They also clear non-argument registers before entering the callee and non-result registers before returning, using function-signature metadata from both parties to determine the permitted interface (see \Cref{sec:reg-confidentiality}).
    These measures prevent a malicious callee from corrupting preserved registers and prevent either party from recovering residual register contents, including leakage caused by mismatched function declarations.

    \item[Control Flow Integrity]
    Inter-compartment calls and returns are tracked in the Trusted Stack and Stack Lookup Table (see \Cref{sec:stack-safety}), and returning trampoline legs use this state alone to identify and restore the caller (see \Cref{sec:cfi}).
    These invariants neutralize attempts by a malicious callee to reuse a retained return pointer, substitute another trampoline's returning leg, or otherwise return into a compartment under the wrong calling context.
\end{description}

\section{Evaluation}\label{sec:eval}

We first present case studies that demonstrate particular aspects of our design, including sub-libraries, debugging, tracing, and managed language support.
We then present benchmark results involving the platforms listed in \cref{tab:eval-platforms}, measured on non-debug builds of CheriBSD\footnote{On the Morello platform, we use the Benchmark ABI~\cite{watson_early_2023} to work around micro-architectural quirks and simulate performance achievable by an ideal implementation.} with global heap temporal safety~\cite{filardo_cornucopia_2024} enabled\footnote{Global heap temporal safety is disabled for the gRPC benchmark. See \Cref{sec:grpc}.}.
\begin{table}[h]
\centering
\caption{Evaluated CHERI-Extended Platforms}\label{tab:eval-platforms}
\begin{compactcontents}
\begin{tabularx}{\columnwidth}{Y l >{\hsize=0.8\hsize}Y >{\hsize=1.2\hsize}Y}
    \toprule
    \textbf{Processor} & \textbf{Base ISA} & \textbf{$\mu$Architecture} & \textbf{Remark} \\
    \midrule
    Arm Morello \newline (ASIC) & Armv8-A & Superscalar, Out-of-Order & Research design~\cite{grisenthwaite_arm_2023} \\
    \midrule
    Cambridge \newline Toooba (FPGA) & RISC-V & Superscalar, Out-of-Order & Research design~\cite{rugg_suite_2023} \\
    \midrule
    Codasip X730 \newline (FPGA) & RISC-V & Dual-issue, 9-stage & Industrial\newline proprietary design \\
    \midrule
    CapLtd CVA6-\newline CHERI (FPGA) & RISC-V & Dual-issue, 6-stage & Industrial open-\newline source design~\cite{cva6_capltd_2026} \\
    \bottomrule
\end{tabularx}
\end{compactcontents}
\end{table}

\subsection{Case Study: tcpdump}\label{sec:tcpdump}

Tcpdump is a network analysis tool that processes packets captured in real time or from a saved file.
It uses a chain of dissectors, each printing a human-readable decoding of the header before passing the remaining data to the next relevant dissector.
This processing of untrustworthy data is often performed in a single process as a highly privileged user, making tcpdump a natural candidate for compartmentalization.\footnote{On some platforms, tcpdump supports privilege separation, but this is not universal.}
The tcpdump build system already places dissectors in an internal library, \texttt{libnetdissect}, which we first extend to a shared library.
We then use a mix of generated and hand-written policies (the latter shown below) to further split it into infrastructure and per-protocol sub-libraries following existing software engineering boundaries: each \texttt{print-<proto>.c} is placed in its own compartment; NFS and SMB protocol helper functions, as well as shared cryptographic signature validation routines, are placed in their own respective compartments; finally, utility functions such as \texttt{txtproto\_print} end up in the \texttt{netdissect} compartment, a catch-all compartment described in \Cref{sec:applying-placement-rules}.
These simple and maintainable build system changes, made in a few hours, dramatically scale tcpdump from 11 compartments to 180.

\begin{verbatimcontents}
\begin{verbatim}
    {"compartments": {
      "print-nfs": {"files": ["parsenfsfh.*o"]},
      "print-smb": {"files": ["smbutil.*o"]},
      "signature": {"files": ["signature.*o"]}}}
\end{verbatim}
\end{verbatimcontents}

Our model's minimal perturbation to the ABI allows debuggers to be extended with compartment awareness.
\Cref{fig:tcpdump-stack-trace} shows a stack trace, generated by an extended version of GDB, of tcpdump when printing an HTTP packet.
Similar to the contiguous abstract stack of \Cref{sec:stack-safety}, call frames shown in black are interspersed with colored trusted frames that record compartment transitions, which mirror the Ethernet-IP-TCP-HTTP packet hierarchy.\footnote{The transition through the TCP dissector compartment is absent because \texttt{tcp\_print} tail-calls \texttt{http\_print}.
As explained in \Cref{sec:tail-calls}, this causes the topmost trusted frame to be overwritten with the HTTP dissector compartment as the callee.}

\begin{figure}[h]
\centering
\begin{verbatimcontents}
\input{graphs/tcpdump-stack-trace}
\end{verbatimcontents}
\caption{tcpdump stack trace when printing an HTTP packet.}\label{fig:tcpdump-stack-trace}
\end{figure}

\begin{figure*}[t]
\centering
  \includegraphics{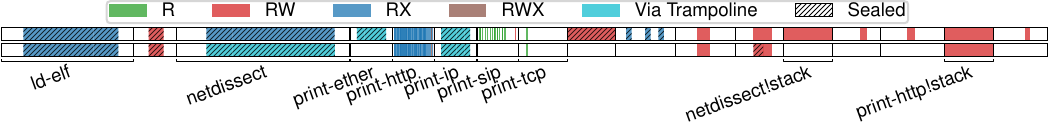}
\caption{Memory reachable from \texttt{http\_print} before and after sealing access to \texttt{jmp\_buf}. Labeled segments include sub-libraries and stacks, and unlabeled ones include heaps and other custom mappings. Hatched intervals are covered \emph{only} by sealed capabilities, and cyan intervals are reachable \emph{only} through calling or returning via trampolines. Small intervals are widened for visibility. \texttt{RWX} intervals displayed here hold PC-reachable mutable data, which are not actually mapped on executable pages.}\label{fig:tcpdump}
\end{figure*}

The policies define compartment boundaries, but validating their security benefit requires understanding how each compartment's privileges evolve at run time as capabilities are propagated.
As a first step in this direction, we present \emph{CheriTree}, a tool that traverses the capability graph (see \Cref{sec:cheri-primer}) at any given point in execution to map out reachable memory as intervals on the address-space ``spectrum'' colored by the union of the permissions of all capabilities covering each interval.
This traversal is exact because all valid capabilities can be precisely identified in memory by their tags, and their bounds enable an exhaustive graph search.

We first take a snapshot of tcpdump compartmentalized with \emph{no policy} in \texttt{http\_print}.
Reachable memory totals 8.37\,MB, with 4.78\,MB being read-writable.
With the policy, the numbers drop to 814\,KB and 408\,KB, and the address space is visualized in the upper row of \Cref{fig:tcpdump}, with over 100 unreachable compartments and memory mappings omitted.
As expected, apart from \texttt{print-http} itself and its stack, the dynamic linker (\texttt{ld-elf}) and the caller (\texttt{print-ip}) are reachable, gated by sentries and trampolines, respectively.
\emph{Unexpectedly}, \texttt{print-ether} and the entire stack of the catch-all \texttt{netdissect} compartment are also reachable.
Examining the memory with GDB reveals this to be due to a \texttt{jmp\_buf} reachable from a function argument.
And because this argument is used as an opaque pointer, a simple refactoring can seal it into a sentry (see \Cref{sec:cheri-primer}), which is non-dereferenceable.
Indeed, this closes both gaps and a few more, as shown in the lower plot.
Total reachable memory further falls to 527\,KB, with 150\,KB being read-writable.

To measure the performance, we first use tcpdump compartmentalized with \emph{no policy} to print a capture file containing 100K packets of various protocols at maximum verbosity on Morello.
Compartment transition tracing shows that there are on average 384 transitions per packet.
Tcpdump compartmentalized with the policy is then run with the same input.
Tracing shows that only 34 compartments out of 180 are entered at least once, and 659 compartment transitions occur per packet---a 71.6\% increase.
The CPU cycle count increases by 33.3\% from 2.84B to 3.79B.

\subsection{Case Study: Chromium and V8}\label{sec:v8}

Chromium is an open-source web browser underpinning widely used applications such as Google Chrome and VS Code.
We take an existing research prototype port of Chromium to CheriABI for Morello, which consists of 47M LoC, and adapt it to work with linkage-based compartmentalization.
The adaptation focuses on the bundled V8 JavaScript engine---a complex managed language runtime containing multiple JIT compilers and garbage collectors---and changes fewer than 300 lines out of V8's roughly 2M LoC ($<$0.015\%).
By successfully adapting V8 with limited effort, we demonstrate the generality of our approach in supporting diverse language features.

\paragraph{Garbage Collection}
Some of V8's garbage collectors periodically scan the callee-saved registers and the stack to mark live objects, as illustrated in \Cref{alg:stack-scan-before}.
The core stack scanning loop is adapted for linkage-based compartmentalization using the same APIs introduced for \texttt{setjmp}/\texttt{longjmp} and C++ exceptions in \Cref{sec:setjmp} to traverse disjoint stacks belonging to different compartments, resulting in \Cref{alg:stack-scan-after}.
\begin{algorithm}
\caption{Original V8 Core Stack Scanning Loop}\label{alg:stack-scan-before}
\begin{compactcontents}
\begin{algorithmic}[1]
\ForAll{$r \in \operatorname{GetCalleeSaved}()$}
    \State Push $r$ onto the stack
\EndFor
\State $begin \gets \operatorname{GetStackBegin}()$
\State $end \gets \operatorname{GetStackEnd}()$
\For{$p \gets begin$ \textbf{to} $end$ \textbf{step} $16$}
    \State \Call{Visit}{$*p$}
\EndFor
\end{algorithmic}
\end{compactcontents}
\end{algorithm}
\begin{algorithm}
\caption{Adapted V8 Core Stack Scanning Loop}\label{alg:stack-scan-after}
\begin{compactcontents}
\begin{algorithmic}[1]
\State $f \gets \operatorname{GetTopTrustedFrame}()$
\Repeat
    \For{$p \gets f.\mathrm{stack\_begin}$ \textbf{to} $f.\mathrm{stack\_end}$ \textbf{step} $16$}
        \State \Call{Visit}{$*p$}
    \EndFor
    \ForAll{$r \in f.\mathrm{callee\_saved}$}
        \State \Call{Visit}{$r$}
    \EndFor
    \State $f \gets \operatorname{GetNextTrustedFrame}(f)$
\Until{$f = \mathrm{null}$}
\end{algorithmic}
\end{compactcontents}
\end{algorithm}

\paragraph{Built-Ins}
Through a complex process, V8's build system pre-compiles JavaScript's built-in functions from a DSL into optimized native code snippets that JIT-compiled code can invoke at run time.
These code snippets are then packed into a blob of assembly instructions to be embedded in the executable section of the V8 binary, even though they are treated as mere data to perform the equivalent of dynamic linking.
To prevent V8's pointer to this blob from being treated as a function pointer and wrapped in a trampoline, which would render it unusable as a data pointer, we relocate it using the ``code pointer'' relocation described in \Cref{sec:setjmp}.

\paragraph{Outcome}
With compartmentalization enabled, a \emph{component} build of Chromium, which does not statically link most dependent libraries into the main executable, obtains 536 run-time compartments with no policy.
V8's \textasciitilde{}115K tests, which heavily exercise garbage collection and JIT compilation, pass at the same rate of \textasciitilde{}98\% ($\pm$0.5\% due to test non-determinism) as the baseline.

\subsection{Micro-benchmark: Compartment Transition}\label{sec:microbench}

We use performance counters to analyze the best-case overhead of a compartment transition.
The experiment makes a large number of calls to a function in another library that returns immediately.
The call goes through both the PLT and a trampoline when compartmentalization is enabled and only the PLT when it is disabled.
Counter values for the entire run are divided by the number of calls to compute the cost of a single call and amortize the probe effect.

\begin{figure}[h]
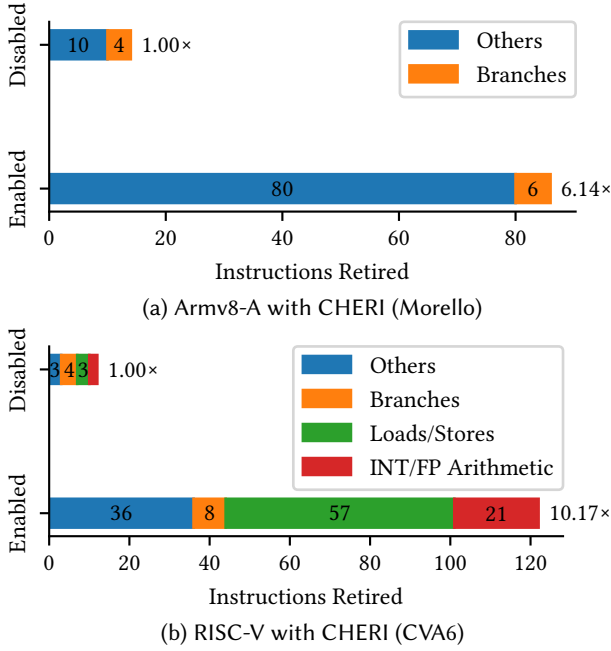

\subfloat[Armv8-A with CHERI (Morello)]{\input{plots/purecap_benchmark-func_nowork-0-False.pgf}}
\hfil
\centering
\subfloat[RISC-V with CHERI (CVA6)]{\input{plots/cva6-purecap-func_nowork-0-False.pgf}}
\caption{Instruction type breakdown of an inter-library function call with and without compartmentalization, measured for Armv8-A and RISC-V. Only branch instruction retirement counters are available on Morello. Pointer arithmetic instructions are classified as ``Others'' on CVA6.}\label{fig:breakdown}
\end{figure}

\begin{table}[h]
\centering
\caption{Cycle Count of an Inter-library Function Call under Different Compartmentalization Settings}\label{tab:cycle-overhead}
\begin{compactcontents}
\begin{tabular}{lllrrr}
  \toprule
  \textbf{Processor} & \textbf{Base ISA} & \textbf{OoO} & \textbf{Disabled} & \textbf{Enabled} & \textbf{Overhead} \\
  \midrule
  Morello & Armv8-A & Yes & 10 & 60 & 6.0$\times$ \\
  Toooba  & RISC-V  & Yes & 8 & 96 & 12.0$\times$ \\
  X730    & RISC-V  & No  & 10 & 121 & 12.1$\times$ \\
  CVA6    & RISC-V  & No  & 20 & 155 & 7.8$\times$ \\
  \bottomrule
\end{tabular}

\end{compactcontents}
\end{table}

Armv8-A's conditional instructions enable trampolines to use fewer branch instructions than on RISC-V, as shown in \Cref{fig:breakdown}.

Micro-architecture considerably affects the CPU cycle overhead introduced by trampolines.
As shown in \Cref{tab:cycle-overhead}, out-of-order processors consume fewer cycles regardless of whether compartmentalization is enabled.
Between in-order processors, the proprietary X730 dominates the less optimized CVA6 in absolute terms, but the poorer baseline of the latter results in superior relative overhead.

\subsection{Macro-benchmark: FFmpeg Decoding}\label{sec:ffmpeg}

Media codecs are widely deployed but are highly susceptible to malicious input.
We use performance counters to analyze the overhead of compartmentalizing \texttt{libdav1d}, a popular decoder for the mainstream AV1 format. 
The experiment on Morello uses FFmpeg to decode 10-second videos at three resolutions, five times each.
As shown in \Cref{tab:dav1d}, up to 1.15 million compartment transitions incur only small micro-architectural overheads.

\begin{table}[b]
\centering
\caption{Relative Overhead of Decoding 10-Second Videos}\label{tab:dav1d}
\begin{compactcontents}

\begin{tabular}{lrrrrr}
  \toprule
  \textbf{Resolution} & \textbf{Transition} & \textbf{$\Delta$CPU} & \textbf{$\Delta$Inst.} & \textbf{$\Delta$Branch} & \textbf{$\Delta$Mem.} \\
   & \textbf{Count} & \textbf{Cycles} & \textbf{Retired} & \textbf{Retired} & \textbf{Access} \\
  \midrule
  $1280 \times 720$  & 751K  & 0.92\% & 0.53\% & 1.50\% & 0.71\% \\
  $1920 \times 1080$ & 849K  & 0.68\% & 0.49\% & 1.29\% & 0.60\% \\
  $3840 \times 2160$ & 1.15M & 0.29\% & 0.30\% & 0.99\% & 0.40\% \\
  \bottomrule
\end{tabular}

\end{compactcontents}
\end{table}

\subsection{Macro-benchmark: SPDK}\label{sec:spdk}

\begin{figure}[t]
\centering
\input{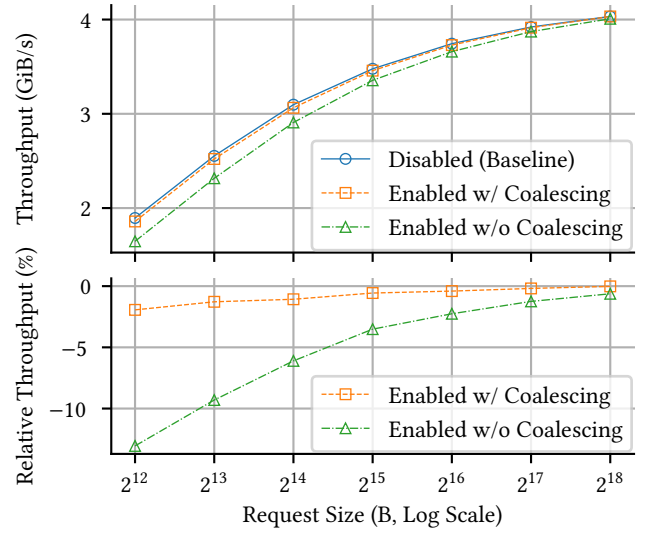}
\caption{SPDK throughput under different compartmentalization settings.}\label{fig:spdk-overhead}
\end{figure}

Storage Performance Development Kit (SPDK)~\cite{yang_spdk_2017} is a user-space storage stack widely deployed in enterprise storage products and cloud-native solutions.
It depends on multiple independent projects such as Data Plane Development Kit (DPDK), totaling over 4M lines of code in over 200 libraries.
The libraries, although intended for functional separation rather than security boundaries, provide a valuable platform for performance evaluation.

SPDK 22.11 is modified to remove the dependency on the DPDK contiguous memory kernel driver for ease of benchmarking.
Performance is determined to be equivalent.
The experiment on Morello uses \texttt{fio} 3.41 with the SPDK plugin to generate random reads with various request sizes against a RAM disk.
The baseline disables compartmentalization, and two other configurations are measured:
\begin{enumerate}
    \item SPDK and DPDK libraries are coalesced into 2 separate compartments, totaling 28 run-time compartments.
    \item All libraries are separate compartments, increasing the number of run-time compartments to 78.
\end{enumerate}

As shown in \Cref{fig:spdk-overhead}, throughput for 4\,KB reqeusts decreases by 1.9\% when coalescing compartments and by 13\% without coalescing.
The impact decreases as the request size increases.
To model a worst-case scenario of throughput reduction, SPDK is then modified to transfer only a single byte at a time rather than entire data blocks.
The baseline rate is 1209K IOPS, and compartmentalization reduces it by 4.4\% and 26\% with and without coalescing, respectively.

\subsection{Macro-benchmark: gRPC}\label{sec:grpc}

gRPC is a high-performance C/C++ remote procedure call framework deployed across data center services and edge devices.
It has the Abseil C++ library as a notable dependency.

\begin{figure}[b]
\centering
\input{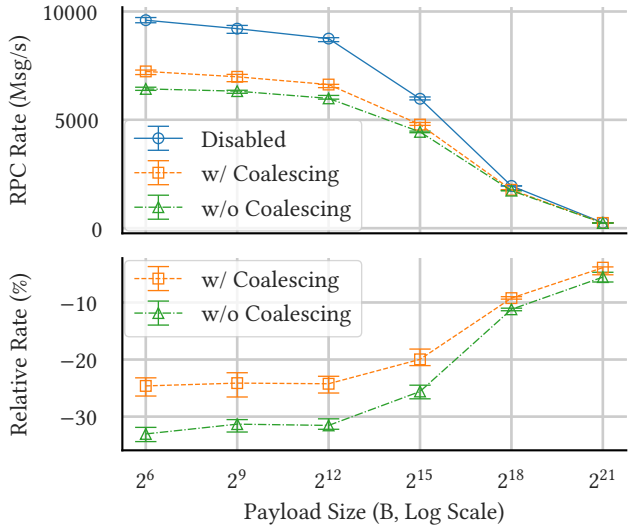}
\caption{gRPC median message rate under different compartmentalization settings with 95\% confidence intervals.}\label{fig:grpc-overhead}
\end{figure}

The experiment on Morello runs the queries-per-second (QPS) benchmark from gRPC 1.54.2.
A server and client are pinned to different cores in the same cluster and communicate through the loopback interface.
The workload exchanges unary RPC calls over a single channel in ``synchronous'' mode.
The client maintains 8 outstanding RPC requests (one per thread) and disables TLS.
For each payload size, 20 samples are collected, each lasting 60\,s after a 15\,s warm-up period.
The baseline disables compartmentalization, and two compartmentalization settings are tested:
\begin{enumerate}
    \item Core gRPC, Abseil, and cryptographic libraries are coalesced into 3 compartments, totaling 18 run-time compartments.
    \item All libraries are separate compartments, increasing the number of run-time compartments to 82.
\end{enumerate}
For this benchmark, global heap temporal safety is disabled (cf. \Cref{sec:eval}) due to unexpected, workload-specific performance penalties when it is enabled together with compartmentalization, an effect that appears to be addressable through further engineering.

This benchmark highlights the importance of the fast paths in trampolines described in \Cref{sec:trampoline}.
In a sample run with compartment coalescing and 64\,B payloads before deploying fast paths, out of a total of 30.4M compartment transitions, 19.6M (64\%) are compartment \emph{self-transitions} almost entirely due to calling C++ virtual functions defined in the same compartment.
Deploying fast paths dramatically cuts the number of self-transitions by 99.2\%, reducing the total number of transitions to 10.5M.

As shown in \Cref{fig:grpc-overhead}, without compartment coalescing, for a payload size of 64\,B, the RPC rate decreases by 33\%.
The impact drops to below 5.6\% as the payload size increases.
Coalescing improves this figure to 25\% and 3.9\% for 64\,B and 2\,MB payloads.

\section{Future Work}\label{sec:future}

A recent SoK~\cite{lefeuvre_sok_2025} argues that compartmentalization systems should be evaluated holistically across policy definition, abstractions, and enforcement mechanisms.
Our main contribution lies in enforcement mechanisms, providing a novel level of \emph{fine-grained} and \emph{scalable} enforcement of compartmentalization guarantees such as ABI hygiene.
A 2020 Microsoft study~\cite{joly_security_2020} estimated that a CHERI system would deterministically mitigate at least 67\% of then-reported memory safety vulnerabilities, leaving substantial residual risks.
JIT compilers also present a significant attack surface where code-generation flaws can produce unsafe code even when the compiler itself is memory safe.
Further limiting attackers' abilities therefore requires techniques described in the present work and beyond:

\begin{description}
\item[Interface Safety]
Although ABI hygiene minimizes the unintentional sharing of objects between compartments, programmers remain responsible for safeguarding against compartment interface vulnerabilities~\cite{lefeuvre_assessing_2023} that arise from intentional sharing.
Tools like RLBox~\cite{narayan_retrofitting_2020} provide type-driven taint tracking and mandatory data validation to harden compartment interfaces.
Our model could serve as an efficient backend for such tools by replacing many software-enforced properties with hardware-enforced guarantees.

\item[Policy Automation]
Realizing the full benefit of scaling to thousands of compartments requires complementary research into automation, which has evolved from refactoring-oriented tools such as Privtrans~\cite{brumley_privtrans_2004}, Wedge~\cite{bittau_wedge_2008}, and SOAAP~\cite{gudka_clean_2015} to more recent work with tunable security/performance objectives such as SCALPEL~\cite{roessler_scalpel_2021} and HAKC~\cite{mckee_hakc_2022}.
Preliminary results from joint work on these techniques indicate that automated compartmentalization is feasible for libraries at least the scale of \texttt{libpng}.
Ideally, policy construction should also be paired with effective \emph{auditing}.
A useful example is CHERIoT's policy auditing tool~\cite{amar_cheriot_2025}, which allows users to write assertions that are checked against binary images.

\item[Stack Temporal Safety]
Stack references passed across compartment boundaries, including both explicit references to local variables and implicitly allocated buffers for function arguments and return values, can be hoarded beyond their lifetimes~\cite{anderson_formalizing_2023}.
Allocating such ``unsafe'' buffers on a shadow stack~\cite{bhatkar_efficient_2005,kuznetsov_code_2014} partially mitigates this, while proposed hardware-software extensions such as the ``slinky stack'' on CheriOS~\cite{esswood_cherios_2021} and various new kinds of capabilities~\cite{skorstengaard_stktokens_2019,tsampas_temporal_2019,georges_efficient_2021,georges_temps_2022} promise stronger guarantees.

\item[Side-Channel Attacks]
Speculative execution attacks can leak compartment secrets~\cite{fuchs_safe_2024}.
Architectural speculation contracts and micro-architectural identification of compartments have been proposed for CHERI implementations to mitigate such risks~\cite{fuchs_toward_2025}.
However, these remain experimental solutions pending further research.
\end{description}

\section{Related Work}\label{sec:related-work}

\begin{description}
\item[\texttt{libcheri}] 
A prior model~\cite{watson_cheri_2015} uses CHERI to construct sandboxes similar to classes in object-oriented programming.
In contrast, our model 
\begin{enumerate*}[label=\alph*), itemjoin={{, }}, itemjoin*={{, and }}]
    \item is a refinement of existing dynamic linking rather than a standalone runtime
    \item performs compartment transitions in user-space without trapping into the kernel
    \item supports compartment transitions via function pointers, enabling callbacks and C++ virtual functions
    \item handles complex language features such as JIT compilation and garbage collection
    \item demonstrates integration with debugging and visualization tools
    \item is accompanied by an open-source software evaluation corpus of over 50M LoC of C/C++
    \item works on multiple processors, including Arm's Morello
\end{enumerate*}.

\item[Embedded CHERI]
CheriRTOS~\cite{xia_cherirtos_2018} introduces CHERI to embedded systems, and CompartOS~\cite{almatary_compartos_2022} implements linkage-based compartmentalization with fault tolerance.
CHERIoT~\cite{amar_cheriot_2025} is a commercial hardware-software platform that additionally provides stack and heap temporal safety, a clean-slate OS practicing fine-grained privilege delegation and minimization, and powerful policy auditing facilities.
Our work is inspired by these designs but the challenges of targeting a general-purpose system necessitate trade-offs:
\begin{itemize}
    \item Embedded systems can afford to impose clean-slate resource allocation, privilege delegation, synchronization, and fault tolerance abstractions on applications, whereas our design currently aims not to cause such disruptions, limiting our support to POSIX.
    \item CHERIoT's stack temporal safety enforcement forbids global and heap memory from storing stack references, which enables the system to limit their lifetime simply through stack scanning.
    However, this restriction is incompatible with the C standard and indeed much existing code, while stack scanning is too costly on general-purpose systems with multi-megabyte stacks.
    Thus, we defer addressing stack temporal safety to future work.
    \item Compared to CHERIoT where the integrator is solely responsible for auditing the policy of a static system, our design faces multiple parties---developers, administrators, and users---who jointly and dynamically interact with, and even alter, the system.
    The desired \emph{distributed} policy auditing remains an open problem.
\end{itemize}

\item[Formal Foundations]
Formal modeling of software on capability systems complements our work by proving similar security properties under idealized models.
CapablePtrs~\cite{korashy_capableptrs_2021} proves that pointer-as-capabilities compilation does not increase the attack surface beyond a unit's source-level interface, assuming a TCB close to ours.
Skorstengaard et al.~\cite{skorstengaard_reasoning_2019} present an alternative capability-based calling convention with proven stack isolation and well-bracketed control-flow properties, followed by successive proposals with improved efficiency through architectural extensions~\cite{skorstengaard_stktokens_2019,georges_efficient_2021,georges_temps_2022}.

\item[Memory Tagging]
MTE~\cite{arm_memory_2019} supports 16 colors at pointer-sized granularity, which helps protect against buffer overflows and use-after-free vulnerabilities.
However, the small number of domains is insufficient to enable compartmentalization.
LatticeBox~\cite{liu_latticebox_2026} represents compartment relations as hierarchical labels, compactly encoding memory accessibility without access-control tables.
However, this accessibility relation is still pre-configured, so dynamic, data-dependent memory sharing remains a challenge.
PUMP~\cite{dhawan_architectural_2015} and PIPE~\cite{sullivan_dover_2017} are significantly more powerful tagging schemes that allow for not only larger tags but also more expressive rules that can automatically propagate tags.
This enables much richer security policies to be implemented, such as control flow integrity (CFI) and fine-grained compartmentalization~\cite{roessler_scalpel_2021}.
However, the centralized rules face a security-performance trade-off when modeling the complex delegation of pointers in modern applications.
In contrast, CHERI's capability-based decentralized enforcement naturally scales for arbitrary pointer flows.

\item[Page-Based Protection]
ERIM~\cite{vahldiek_erim_2019}, Hodor~\cite{hedayati_hodor_2019}, Cerberus~\cite{voulimeneas_cerberus_2022}, Jenny~\cite{schrammel_jenny_2022}, and Endokernel~\cite{yang_endokernel_2024} combine Intel MPK with software safeguards for efficient in-process isolation, but MPK's 16 hardware keys restrict them to a small number of concurrently available protection domains.
Proposed architectural extensions relax this limit but retain other constraints: Donky~\cite{schrammel_donky_2020} expands the key space to 1024 but makes only four keys usable at a time; SecureCells~\cite{bhattacharyya_securecells_2023} supports many more domains, but its centralized permission table operates at VMA granularity and must be updated whenever access is delegated; HFI~\cite{narayan_hfi_2023} supports arbitrarily many sandboxes by reconfiguring region registers on entry, but each active sandbox can access only ten regions.
These architecture-driven constraints considerably restrict the applicability and scalability of such systems in general-purpose software, whose compartments exhibit diverse memory-access and communication patterns.

\item[Software-Based Fault Isolation]
SFI~\cite{wahbe_efficient_1993} restricts memory access by untrusted code within the same process through a combination of compiler instrumentation, static analysis, and dynamic checks. 
Systems like Native Client~\cite{yee_native_2009} and WebAssembly~\cite{haas_bringing_2017} rely on software bounds checking with non-trivial performance overhead. 
Lightweight Fault Isolation (LFI)~\cite{yedidia_lfi_2024} reduces this cost by restricting AArch64 machine code to a stylized, efficiently verifiable form.
These systems offer at most coarse-grained memory sharing and cannot readily share allocations such as stack objects, limiting their use for fine-grained compartmentalization of existing software.
\end{description}

\section{Conclusion}

Linkage-based compartmentalization leverages CHERI memory safety to enable scalable privilege separation within a UNIX process. 
With CheriBSD's system-wide compartmentalization enabled (see \Cref{sec:mech}), a freshly booted Morello system logged into the KDE desktop environment already hosts more than 2000 compartments across all processes.
Launching a moderate number of additional applications causes this number to exceed 10K.
All this is achieved with little, if any, source-code disruption to the applications, and the impact on performance or behavior is imperceptible in our day-to-day software development and web-serving workloads.

Moreover, case studies demonstrate that the design integrates well with realistic software development workflows through its native compiler toolchain, debugger, and tracing support.
Performance evaluations across diverse CHERI-extended processors also confirm the feasibility of this single-address-space approach.

\begin{acks}

We are grateful for assistance and feedback from
David Chisnall,
Alexandre Joannou,
Timothy M. Jones,
Marno van der Maas,
Peter G. Neumann,
George V. Neville-Neil,
Murali Vijayaraghavan, and
Jonathan Woodruff,
as well as additional support from Arm, Codasip, and Google.

Distribution Statement A: Approved for public release; distribution is unlimited.
This work is sponsored in part by the Defense Advanced Research Projects Agency (DARPA) and the Air Force Research Laboratory (AFRL) under contracts
HR0011-22-C-0110 (``ETC''),
HR0011-23-C-0031 (``MTSS''), and
FA8750-24-C-B047 (``DEC'') as part of the DARPA CPM research program, and by the Office of Naval Research (ONR) under contract N00014-22-1-2463 (``SWISH'').
The views, opinions, and/or findings contained in this report are those of the authors and should not be interpreted as representing the official views or policies, either expressed or implied, of the Department of Defense or the U.S. Government.

This work is also sponsored in part by
Innovate UK projects 105694 (``Digital Security by Design (DSbD) Technology Platform Prototype''),
10027440 (``Developing and Evaluating an Open-Source Desktop for Arm Morello''), and
10027332 (``MOJO – A Robust Java Virtual Machine for Morello''), as well as
the EPSRC CHaOS grant (EP/V000292/1),
the EPSRC ``UKRI3001: CHERI Research Centre'' grant,
and a Croucher Scholarship from the Croucher Foundation.

\end{acks}

\bibliographystyle{ACM-Reference-Format}
\bibliography{references,zotero}


\begin{thebibliography}{58}

%%% the Web address.

\ifx \showCODEN    \undefined \def \showCODEN     #1{\unskip}     \fi
\ifx \showISBNx    \undefined \def \showISBNx     #1{\unskip}     \fi
\ifx \showISBNxiii \undefined \def \showISBNxiii  #1{\unskip}     \fi
\ifx \showISSN     \undefined \def \showISSN      #1{\unskip}     \fi
\ifx \showLCCN     \undefined \def \showLCCN      #1{\unskip}     \fi
\ifx \shownote     \undefined \def \shownote      #1{#1}          \fi
\ifx \showarticletitle \undefined \def \showarticletitle #1{#1}   \fi
\ifx \showURL      \undefined \def \showURL       {\relax}        \fi
% invisible to TeX
\providecommand\bibfield[2]{#2}
\providecommand\bibinfo[2]{#2}
\providecommand\natexlab[1]{#1}
\providecommand\showeprint[2][]{arXiv:#2}

\bibitem[Almatary et~al\mbox{.}(2022)]%
        {almatary_compartos_2022}
\bibfield{author}{\bibinfo{person}{Hesham Almatary}, \bibinfo{person}{Michael
  Dodson}, \bibinfo{person}{Jessica Clarke}, \bibinfo{person}{Peter Rugg},
  \bibinfo{person}{Ivan Gomes}, \bibinfo{person}{Michal Podhradsky},
  \bibinfo{person}{Peter~G. Neumann}, \bibinfo{person}{Simon~W. Moore}, {and}
  \bibinfo{person}{Robert N.~M. Watson}.} \bibinfo{year}{2022}\natexlab{}.
\newblock \bibinfo{title}{CompartOS: CHERI Compartmentalization for Embedded
  Systems}.
\newblock
\showeprint[arxiv]{2206.02852}~[cs.CR]
\urldef\tempurl%
\url{https://arxiv.org/abs/2206.02852}
\showURL{%
\tempurl}


\bibitem[Amar et~al\mbox{.}(2025)]%
        {amar_cheriot_2025}
\bibfield{author}{\bibinfo{person}{Saar Amar}, \bibinfo{person}{Tony Chen},
  \bibinfo{person}{David Chisnall}, \bibinfo{person}{Nathaniel~Wesley Filardo},
  \bibinfo{person}{Ben Laurie}, \bibinfo{person}{Hugo Lefeuvre},
  \bibinfo{person}{Kunyan Liu}, \bibinfo{person}{Simon~W. Moore},
  \bibinfo{person}{Robert Norton-Wright}, \bibinfo{person}{Margo Seltzer},
  \bibinfo{person}{Yucong Tao}, \bibinfo{person}{Robert N.~M. Watson}, {and}
  \bibinfo{person}{Hongyan Xia}.} \bibinfo{year}{2025}\natexlab{}.
\newblock \showarticletitle{CHERIoT RTOS: An OS for Fine-Grained Memory-Safe
  Compartments on Low-Cost Embedded Devices}. In
  \bibinfo{booktitle}{\emph{Proceedings of the ACM SIGOPS 31st Symposium on
  Operating Systems Principles}} (Lotte Hotel World, Seoul, Republic of Korea)
  \emph{(\bibinfo{series}{SOSP '25})}. \bibinfo{publisher}{Association for
  Computing Machinery}, \bibinfo{address}{New York, NY, USA},
  \bibinfo{pages}{67–84}.
\newblock
\showISBNx{9798400718700}
\href{https://doi.org/10.1145/3731569.3764844}{doi:\nolinkurl{10.1145/3731569.3764844}}


\bibitem[Anderson et~al\mbox{.}(2023)]%
        {anderson_formalizing_2023}
\bibfield{author}{\bibinfo{person}{Sean~Noble Anderson},
  \bibinfo{person}{Roberto Blanco}, \bibinfo{person}{Leonidas Lampropoulos},
  \bibinfo{person}{Benjamin~C. Pierce}, {and} \bibinfo{person}{Andrew
  Tolmach}.} \bibinfo{year}{2023}\natexlab{}.
\newblock \showarticletitle{Formalizing Stack Safety as a Security Property}.
  In \bibinfo{booktitle}{\emph{2023 IEEE 36th Computer Security Foundations
  Symposium (CSF)}}. \bibinfo{pages}{356--371}.
\newblock
\showISSN{2374-8303}
\href{https://doi.org/10.1109/CSF57540.2023.00037}{doi:\nolinkurl{10.1109/CSF57540.2023.00037}}


\bibitem[{Arm}(2019)]%
        {arm_memory_2019}
\bibfield{author}{\bibinfo{person}{{Arm}}.} \bibinfo{year}{2019}\natexlab{}.
\newblock \bibinfo{title}{Memory {Tagging} {Extension}: {Enhancing} memory
  safety through architecture}.
\newblock
\urldef\tempurl%
\url{https://developer.arm.com/community/arm-community-blogs/b/architectures-and-processors-blog/posts/enhancing-memory-safety}
\showURL{%
\tempurl}


\bibitem[Bauereiss et~al\mbox{.}(2022)]%
        {bauereiss_verified_2022}
\bibfield{author}{\bibinfo{person}{Thomas Bauereiss}, \bibinfo{person}{Brian
  Campbell}, \bibinfo{person}{Thomas Sewell}, \bibinfo{person}{Alasdair
  Armstrong}, \bibinfo{person}{Lawrence Esswood}, \bibinfo{person}{Ian Stark},
  \bibinfo{person}{Graeme Barnes}, \bibinfo{person}{Robert N.~M. Watson}, {and}
  \bibinfo{person}{Peter Sewell}.} \bibinfo{year}{2022}\natexlab{}.
\newblock \showarticletitle{Verified Security for the Morello
  Capability-enhanced Prototype Arm Architecture}. In
  \bibinfo{booktitle}{\emph{Programming Languages and Systems}},
  \bibfield{editor}{\bibinfo{person}{Ilya Sergey}} (Ed.).
  \bibinfo{publisher}{Springer International Publishing},
  \bibinfo{address}{Cham}, \bibinfo{pages}{174--203}.
\newblock
\showISBNx{978-3-030-99336-8}


\bibitem[Bhatkar and DuVarney(2005)]%
        {bhatkar_efficient_2005}
\bibfield{author}{\bibinfo{person}{Sandeep Bhatkar} {and}
  \bibinfo{person}{Daniel~C. DuVarney}.} \bibinfo{year}{2005}\natexlab{}.
\newblock \showarticletitle{Efficient Techniques for Comprehensive Protection
  from Memory Error Exploits}. In \bibinfo{booktitle}{\emph{14th USENIX
  Security Symposium (USENIX Security 05)}}. \bibinfo{publisher}{USENIX
  Association}, \bibinfo{address}{Baltimore, MD}.
\newblock
\urldef\tempurl%
\url{https://www.usenix.org/conference/14th-usenix-security-symposium/efficient-techniques-comprehensive-protection-memory-error}
\showURL{%
\tempurl}


\bibitem[Bhattacharyya et~al\mbox{.}(2023)]%
        {bhattacharyya_securecells_2023}
\bibfield{author}{\bibinfo{person}{Atri Bhattacharyya},
  \bibinfo{person}{Florian Hofhammer}, \bibinfo{person}{Yuanlong Li},
  \bibinfo{person}{Siddharth Gupta}, \bibinfo{person}{Andres Sanchez},
  \bibinfo{person}{Babak Falsafi}, {and} \bibinfo{person}{Mathias Payer}.}
  \bibinfo{year}{2023}\natexlab{}.
\newblock \showarticletitle{{ SecureCells: A Secure Compartmentalized
  Architecture }}. In \bibinfo{booktitle}{\emph{2023 IEEE Symposium on Security
  and Privacy (SP)}}. \bibinfo{publisher}{IEEE Computer Society},
  \bibinfo{address}{Los Alamitos, CA, USA}, \bibinfo{pages}{2921--2939}.
\newblock
\href{https://doi.org/10.1109/SP46215.2023.10179472}{doi:\nolinkurl{10.1109/SP46215.2023.10179472}}


\bibitem[Bittau et~al\mbox{.}(2008)]%
        {bittau_wedge_2008}
\bibfield{author}{\bibinfo{person}{Andrea Bittau}, \bibinfo{person}{Petr
  Marchenko}, \bibinfo{person}{Mark Handley}, {and} \bibinfo{person}{Brad
  Karp}.} \bibinfo{year}{2008}\natexlab{}.
\newblock \showarticletitle{Wedge: Splitting Applications into
  {Reduced-Privilege} Compartments}. In \bibinfo{booktitle}{\emph{5th USENIX
  Symposium on Networked Systems Design and Implementation (NSDI 08)}}.
  \bibinfo{publisher}{USENIX Association}, \bibinfo{address}{San Francisco,
  CA}.
\newblock
\urldef\tempurl%
\url{https://www.usenix.org/conference/nsdi-08/wedge-splitting-applications-reduced-privilege-compartments}
\showURL{%
\tempurl}


\bibitem[Brumley and Song(2004)]%
        {brumley_privtrans_2004}
\bibfield{author}{\bibinfo{person}{David Brumley} {and} \bibinfo{person}{Dawn
  Song}.} \bibinfo{year}{2004}\natexlab{}.
\newblock \showarticletitle{Privtrans: Automatically Partitioning Programs for
  Privilege Separation}. In \bibinfo{booktitle}{\emph{13th USENIX Security
  Symposium (USENIX Security 04)}}. \bibinfo{publisher}{USENIX Association},
  \bibinfo{address}{San Diego, CA}.
\newblock
\urldef\tempurl%
\url{https://www.usenix.org/conference/13th-usenix-security-symposium/privtrans-automatically-partitioning-programs-privilege}
\showURL{%
\tempurl}


\bibitem[Davis et~al\mbox{.}(2019)]%
        {davis_cheriabi_2019}
\bibfield{author}{\bibinfo{person}{Brooks Davis}, \bibinfo{person}{Robert N.~M.
  Watson}, \bibinfo{person}{Alexander Richardson}, \bibinfo{person}{Peter~G.
  Neumann}, \bibinfo{person}{Simon~W. Moore}, \bibinfo{person}{John Baldwin},
  \bibinfo{person}{David Chisnall}, \bibinfo{person}{Jessica Clarke},
  \bibinfo{person}{Nathaniel~Wesley Filardo}, \bibinfo{person}{Khilan Gudka},
  \bibinfo{person}{Alexandre Joannou}, \bibinfo{person}{Ben Laurie},
  \bibinfo{person}{A.~Theodore Markettos}, \bibinfo{person}{J.~Edward Maste},
  \bibinfo{person}{Alfredo Mazzinghi}, \bibinfo{person}{Edward~Tomasz
  Napierala}, \bibinfo{person}{Robert~M. Norton}, \bibinfo{person}{Michael
  Roe}, \bibinfo{person}{Peter Sewell}, \bibinfo{person}{Stacey Son}, {and}
  \bibinfo{person}{Jonathan Woodruff}.} \bibinfo{year}{2019}\natexlab{}.
\newblock \showarticletitle{CheriABI: Enforcing Valid Pointer Provenance and
  Minimizing Pointer Privilege in the POSIX C Run-time Environment}. In
  \bibinfo{booktitle}{\emph{Proceedings of the Twenty-Fourth International
  Conference on Architectural Support for Programming Languages and Operating
  Systems}} (Providence, RI, USA) \emph{(\bibinfo{series}{ASPLOS '19})}.
  \bibinfo{publisher}{Association for Computing Machinery},
  \bibinfo{address}{New York, NY, USA}, \bibinfo{pages}{379–393}.
\newblock
\showISBNx{9781450362405}
\href{https://doi.org/10.1145/3297858.3304042}{doi:\nolinkurl{10.1145/3297858.3304042}}


\bibitem[Dhawan et~al\mbox{.}(2015)]%
        {dhawan_architectural_2015}
\bibfield{author}{\bibinfo{person}{Udit Dhawan}, \bibinfo{person}{Catalin
  Hritcu}, \bibinfo{person}{Raphael Rubin}, \bibinfo{person}{Nikos Vasilakis},
  \bibinfo{person}{Silviu Chiricescu}, \bibinfo{person}{Jonathan~M. Smith},
  \bibinfo{person}{Thomas~F. Knight}, \bibinfo{person}{Benjamin~C. Pierce},
  {and} \bibinfo{person}{Andre DeHon}.} \bibinfo{year}{2015}\natexlab{}.
\newblock \showarticletitle{Architectural Support for Software-Defined Metadata
  Processing}.
\newblock \bibinfo{journal}{\emph{SIGARCH Comput. Archit. News}}
  \bibinfo{volume}{43}, \bibinfo{number}{1} (\bibinfo{date}{March}
  \bibinfo{year}{2015}), \bibinfo{pages}{487–502}.
\newblock
\showISSN{0163-5964}
\href{https://doi.org/10.1145/2786763.2694383}{doi:\nolinkurl{10.1145/2786763.2694383}}


\bibitem[El-Korashy et~al\mbox{.}(2021)]%
        {korashy_capableptrs_2021}
\bibfield{author}{\bibinfo{person}{Akram El-Korashy}, \bibinfo{person}{Stelios
  Tsampas}, \bibinfo{person}{Marco Patrignani}, \bibinfo{person}{Dominique
  Devriese}, \bibinfo{person}{Deepak Garg}, {and} \bibinfo{person}{Frank
  Piessens}.} \bibinfo{year}{2021}\natexlab{}.
\newblock \showarticletitle{CapablePtrs: Securely Compiling Partial Programs
  Using the Pointers-as-Capabilities Principle}. In
  \bibinfo{booktitle}{\emph{2021 IEEE 34th Computer Security Foundations
  Symposium (CSF)}}. \bibinfo{pages}{1--16}.
\newblock
\href{https://doi.org/10.1109/CSF51468.2021.00036}{doi:\nolinkurl{10.1109/CSF51468.2021.00036}}


\bibitem[Esswood(2021)]%
        {esswood_cherios_2021}
\bibfield{author}{\bibinfo{person}{Lawrence~G. Esswood}.}
  \bibinfo{year}{2021}\natexlab{}.
\newblock \bibinfo{booktitle}{\emph{{CheriOS: designing an untrusted
  single-address-space capability operating system utilising capability
  hardware and a minimal hypervisor}}}.
\newblock \bibinfo{type}{{T}echnical {R}eport} UCAM-CL-TR-961.
  \bibinfo{institution}{University of Cambridge, Computer Laboratory}.
\newblock
\href{https://doi.org/10.48456/tr-961}{doi:\nolinkurl{10.48456/tr-961}}


\bibitem[Filardo et~al\mbox{.}(2024)]%
        {filardo_cornucopia_2024}
\bibfield{author}{\bibinfo{person}{Nathaniel~Wesley Filardo},
  \bibinfo{person}{Brett~F. Gutstein}, \bibinfo{person}{Jonathan Woodruff},
  \bibinfo{person}{Jessica Clarke}, \bibinfo{person}{Peter Rugg},
  \bibinfo{person}{Brooks Davis}, \bibinfo{person}{Mark Johnston},
  \bibinfo{person}{Robert Norton}, \bibinfo{person}{David Chisnall},
  \bibinfo{person}{Simon~W. Moore}, \bibinfo{person}{Peter~G. Neumann}, {and}
  \bibinfo{person}{Robert N.~M. Watson}.} \bibinfo{year}{2024}\natexlab{}.
\newblock \showarticletitle{Cornucopia Reloaded: Load Barriers for CHERI Heap
  Temporal Safety}. In \bibinfo{booktitle}{\emph{Proceedings of the 29th ACM
  International Conference on Architectural Support for Programming Languages
  and Operating Systems, Volume 2}} (La Jolla, CA, USA)
  \emph{(\bibinfo{series}{ASPLOS '24})}. \bibinfo{publisher}{Association for
  Computing Machinery}, \bibinfo{address}{New York, NY, USA},
  \bibinfo{pages}{251–268}.
\newblock
\showISBNx{9798400703850}
\href{https://doi.org/10.1145/3620665.3640416}{doi:\nolinkurl{10.1145/3620665.3640416}}


\bibitem[Fuchs(2025)]%
        {fuchs_toward_2025}
\bibfield{author}{\bibinfo{person}{Franz~A. Fuchs}.}
  \bibinfo{year}{2025}\natexlab{}.
\newblock \bibinfo{booktitle}{\emph{{Toward transient-execution attack
  mitigations on CHERI}}}.
\newblock \bibinfo{type}{{T}echnical {R}eport} UCAM-CL-TR-1001.
  \bibinfo{institution}{University of Cambridge, Computer Laboratory}.
\newblock
\href{https://doi.org/10.48456/tr-1001}{doi:\nolinkurl{10.48456/tr-1001}}


\bibitem[Fuchs et~al\mbox{.}(2024)]%
        {fuchs_safe_2024}
\bibfield{author}{\bibinfo{person}{Franz~A. Fuchs}, \bibinfo{person}{Jonathan
  Woodruff}, \bibinfo{person}{Peter Rugg}, \bibinfo{person}{Alexandre Joannou},
  \bibinfo{person}{Jessica Clarke}, \bibinfo{person}{John Baldwin},
  \bibinfo{person}{Brooks Davis}, \bibinfo{person}{Peter~G. Neumann},
  \bibinfo{person}{Robert N.~M. Watson}, {and} \bibinfo{person}{Simon~W.
  Moore}.} \bibinfo{year}{2024}\natexlab{}.
\newblock \showarticletitle{{ Safe Speculation for CHERI }}. In
  \bibinfo{booktitle}{\emph{2024 IEEE 42nd International Conference on Computer
  Design (ICCD)}}. \bibinfo{publisher}{IEEE Computer Society},
  \bibinfo{address}{Los Alamitos, CA, USA}, \bibinfo{pages}{364--372}.
\newblock
\href{https://doi.org/10.1109/ICCD63220.2024.00063}{doi:\nolinkurl{10.1109/ICCD63220.2024.00063}}


\bibitem[Georges et~al\mbox{.}(2021)]%
        {georges_efficient_2021}
\bibfield{author}{\bibinfo{person}{A\"{\i}na~Linn Georges},
  \bibinfo{person}{Arma\"{e}l Gu\'{e}neau}, \bibinfo{person}{Thomas
  Van~Strydonck}, \bibinfo{person}{Amin Timany}, \bibinfo{person}{Alix Trieu},
  \bibinfo{person}{Sander Huyghebaert}, \bibinfo{person}{Dominique Devriese},
  {and} \bibinfo{person}{Lars Birkedal}.} \bibinfo{year}{2021}\natexlab{}.
\newblock \showarticletitle{Efficient and provable local capability revocation
  using uninitialized capabilities}.
\newblock \bibinfo{journal}{\emph{Proc. ACM Program. Lang.}}
  \bibinfo{volume}{5}, \bibinfo{number}{POPL}, Article \bibinfo{articleno}{6}
  (\bibinfo{date}{Jan.} \bibinfo{year}{2021}), \bibinfo{numpages}{30}~pages.
\newblock
\href{https://doi.org/10.1145/3434287}{doi:\nolinkurl{10.1145/3434287}}


\bibitem[Georges et~al\mbox{.}(2022)]%
        {georges_temps_2022}
\bibfield{author}{\bibinfo{person}{A\"{\i}na~Linn Georges},
  \bibinfo{person}{Alix Trieu}, {and} \bibinfo{person}{Lars Birkedal}.}
  \bibinfo{year}{2022}\natexlab{}.
\newblock \showarticletitle{Le temps des cerises: efficient temporal stack
  safety on capability machines using directed capabilities}.
\newblock \bibinfo{journal}{\emph{Proc. ACM Program. Lang.}}
  \bibinfo{volume}{6}, \bibinfo{number}{OOPSLA1}, Article
  \bibinfo{articleno}{74} (\bibinfo{date}{April} \bibinfo{year}{2022}),
  \bibinfo{numpages}{30}~pages.
\newblock
\href{https://doi.org/10.1145/3527318}{doi:\nolinkurl{10.1145/3527318}}


\bibitem[Grisenthwaite et~al\mbox{.}(2023)]%
        {grisenthwaite_arm_2023}
\bibfield{author}{\bibinfo{person}{Richard Grisenthwaite},
  \bibinfo{person}{Graeme Barnes}, \bibinfo{person}{Robert N.~M. Watson},
  \bibinfo{person}{Simon~W. Moore}, \bibinfo{person}{Peter Sewell}, {and}
  \bibinfo{person}{Jonathan Woodruff}.} \bibinfo{year}{2023}\natexlab{}.
\newblock \showarticletitle{The Arm Morello Evaluation Platform—Validating
  CHERI-Based Security in a High-Performance System}.
\newblock \bibinfo{journal}{\emph{IEEE Micro}} \bibinfo{volume}{43},
  \bibinfo{number}{3} (\bibinfo{date}{May} \bibinfo{year}{2023}),
  \bibinfo{pages}{50–57}.
\newblock
\showISSN{0272-1732}
\href{https://doi.org/10.1109/MM.2023.3264676}{doi:\nolinkurl{10.1109/MM.2023.3264676}}


\bibitem[Gudka et~al\mbox{.}(2015)]%
        {gudka_clean_2015}
\bibfield{author}{\bibinfo{person}{Khilan Gudka}, \bibinfo{person}{Robert~N.M.
  Watson}, \bibinfo{person}{Jonathan Anderson}, \bibinfo{person}{David
  Chisnall}, \bibinfo{person}{Brooks Davis}, \bibinfo{person}{Ben Laurie},
  \bibinfo{person}{Ilias Marinos}, \bibinfo{person}{Peter~G. Neumann}, {and}
  \bibinfo{person}{Alex Richardson}.} \bibinfo{year}{2015}\natexlab{}.
\newblock \showarticletitle{Clean Application Compartmentalization with SOAAP}.
  In \bibinfo{booktitle}{\emph{Proceedings of the 22nd ACM SIGSAC Conference on
  Computer and Communications Security}} (Denver, Colorado, USA)
  \emph{(\bibinfo{series}{CCS '15})}. \bibinfo{publisher}{Association for
  Computing Machinery}, \bibinfo{address}{New York, NY, USA},
  \bibinfo{pages}{1016–1031}.
\newblock
\showISBNx{9781450338325}
\href{https://doi.org/10.1145/2810103.2813611}{doi:\nolinkurl{10.1145/2810103.2813611}}


\bibitem[Haas et~al\mbox{.}(2017)]%
        {haas_bringing_2017}
\bibfield{author}{\bibinfo{person}{Andreas Haas}, \bibinfo{person}{Andreas
  Rossberg}, \bibinfo{person}{Derek~L. Schuff}, \bibinfo{person}{Ben~L.
  Titzer}, \bibinfo{person}{Michael Holman}, \bibinfo{person}{Dan Gohman},
  \bibinfo{person}{Luke Wagner}, \bibinfo{person}{Alon Zakai}, {and}
  \bibinfo{person}{JF Bastien}.} \bibinfo{year}{2017}\natexlab{}.
\newblock \showarticletitle{Bringing the web up to speed with WebAssembly}.
\newblock \bibinfo{journal}{\emph{SIGPLAN Not.}} \bibinfo{volume}{52},
  \bibinfo{number}{6}, \bibinfo{pages}{185–200}.
\newblock
\showISSN{0362-1340}
\href{https://doi.org/10.1145/3140587.3062363}{doi:\nolinkurl{10.1145/3140587.3062363}}


\bibitem[Hedayati et~al\mbox{.}(2019)]%
        {hedayati_hodor_2019}
\bibfield{author}{\bibinfo{person}{Mohammad Hedayati},
  \bibinfo{person}{Spyridoula Gravani}, \bibinfo{person}{Ethan Johnson},
  \bibinfo{person}{John Criswell}, \bibinfo{person}{Michael~L. Scott},
  \bibinfo{person}{Kai Shen}, {and} \bibinfo{person}{Mike Marty}.}
  \bibinfo{year}{2019}\natexlab{}.
\newblock \showarticletitle{Hodor: {Intra-Process} Isolation for
  {High-Throughput} Data Plane Libraries}. In \bibinfo{booktitle}{\emph{2019
  USENIX Annual Technical Conference (USENIX ATC 19)}}.
  \bibinfo{publisher}{USENIX Association}, \bibinfo{address}{Renton, WA},
  \bibinfo{pages}{489--504}.
\newblock
\showISBNx{978-1-939133-03-8}
\urldef\tempurl%
\url{http://www.usenix.org/conference/atc19/presentation/hedayati-hodor}
\showURL{%
\tempurl}


\bibitem[Joly et~al\mbox{.}(2020)]%
        {joly_security_2020}
\bibfield{author}{\bibinfo{person}{Nicolas Joly}, \bibinfo{person}{Saif
  ElSherei}, {and} \bibinfo{person}{Saar Amar}.}
  \bibinfo{year}{2020}\natexlab{}.
\newblock \bibinfo{title}{Security analysis of CHERI ISA}.
\newblock
\urldef\tempurl%
\url{https://github.com/microsoft/MSRC-Security-Research/blob/master/papers/2020/Security%20analysis%20of%20CHERI%20ISA.pdf}
\showURL{%
\tempurl}


\bibitem[Kocher et~al\mbox{.}(2019)]%
        {kocher_spectre_2019}
\bibfield{author}{\bibinfo{person}{Paul Kocher}, \bibinfo{person}{Jann Horn},
  \bibinfo{person}{Anders Fogh}, \bibinfo{person}{Daniel Genkin},
  \bibinfo{person}{Daniel Gruss}, \bibinfo{person}{Werner Haas},
  \bibinfo{person}{Mike Hamburg}, \bibinfo{person}{Moritz Lipp},
  \bibinfo{person}{Stefan Mangard}, \bibinfo{person}{Thomas Prescher},
  \bibinfo{person}{Michael Schwarz}, {and} \bibinfo{person}{Yuval Yarom}.}
  \bibinfo{year}{2019}\natexlab{}.
\newblock \showarticletitle{Spectre {Attacks}: {Exploiting} {Speculative}
  {Execution}}. In \bibinfo{booktitle}{\emph{2019 {IEEE} {Symposium} on
  {Security} and {Privacy} ({SP})}}. \bibinfo{publisher}{IEEE},
  \bibinfo{pages}{1--19}.
\newblock
\href{https://doi.org/10.1109/SP.2019.00002}{doi:\nolinkurl{10.1109/SP.2019.00002}}


\bibitem[Kuznetsov et~al\mbox{.}(2014)]%
        {kuznetsov_code_2014}
\bibfield{author}{\bibinfo{person}{Volodymyr Kuznetsov},
  \bibinfo{person}{L\'{a}szl\'{o} Szekeres}, \bibinfo{person}{Mathias Payer},
  \bibinfo{person}{George Candea}, \bibinfo{person}{R. Sekar}, {and}
  \bibinfo{person}{Dawn Song}.} \bibinfo{year}{2014}\natexlab{}.
\newblock \showarticletitle{Code-pointer integrity}. In
  \bibinfo{booktitle}{\emph{Proceedings of the 11th USENIX Conference on
  Operating Systems Design and Implementation}} (Broomfield, CO)
  \emph{(\bibinfo{series}{OSDI'14})}. \bibinfo{publisher}{USENIX Association},
  \bibinfo{address}{USA}, \bibinfo{pages}{147–163}.
\newblock
\showISBNx{9781931971164}


\bibitem[Lefeuvre et~al\mbox{.}(2023)]%
        {lefeuvre_assessing_2023}
\bibfield{author}{\bibinfo{person}{Hugo Lefeuvre}, \bibinfo{person}{Vlad-Andrei
  Bădoiu}, \bibinfo{person}{Yi Chen}, \bibinfo{person}{Felipe Huici},
  \bibinfo{person}{Nathan Dautenhahn}, {and} \bibinfo{person}{Pierre Olivier}.}
  \bibinfo{year}{2023}\natexlab{}.
\newblock \showarticletitle{Assessing the Impact of Interface Vulnerabilities
  in Compartmentalized Software}. In \bibinfo{booktitle}{\emph{Proceedings 2023
  Network and Distributed System Security Symposium}}.
  \bibinfo{publisher}{Internet Society}.
\newblock
\href{https://doi.org/10.14722/ndss.2023.24117}{doi:\nolinkurl{10.14722/ndss.2023.24117}}


\bibitem[Lefeuvre et~al\mbox{.}(2025)]%
        {lefeuvre_sok_2025}
\bibfield{author}{\bibinfo{person}{Hugo Lefeuvre}, \bibinfo{person}{Nathan
  Dautenhahn}, \bibinfo{person}{David Chisnall}, {and} \bibinfo{person}{Pierre
  Olivier}.} \bibinfo{year}{2025}\natexlab{}.
\newblock \showarticletitle{{ SoK: Software Compartmentalization }}. In
  \bibinfo{booktitle}{\emph{2025 IEEE Symposium on Security and Privacy (SP)}}.
  \bibinfo{publisher}{IEEE Computer Society}, \bibinfo{address}{Los Alamitos,
  CA, USA}, \bibinfo{pages}{3107--3126}.
\newblock
\href{https://doi.org/10.1109/SP61157.2025.00075}{doi:\nolinkurl{10.1109/SP61157.2025.00075}}


\bibitem[Limited(2026)]%
        {cva6_capltd_2026}
\bibfield{author}{\bibinfo{person}{Capabilities Limited}.}
  \bibinfo{year}{2026}\natexlab{}.
\newblock \bibinfo{title}{CVA6-CHERI}.
\newblock
\urldef\tempurl%
\url{https://www.capabilitieslimited.co.uk/current-projects/cheri-cva6}
\showURL{%
\tempurl}


\bibitem[Lipp et~al\mbox{.}(2018)]%
        {lipp_meltdown_2018}
\bibfield{author}{\bibinfo{person}{Moritz Lipp}, \bibinfo{person}{Michael
  Schwarz}, \bibinfo{person}{Daniel Gruss}, \bibinfo{person}{Thomas Prescher},
  \bibinfo{person}{Werner Haas}, \bibinfo{person}{Anders Fogh},
  \bibinfo{person}{Jann Horn}, \bibinfo{person}{Stefan Mangard},
  \bibinfo{person}{Paul Kocher}, \bibinfo{person}{Daniel Genkin},
  \bibinfo{person}{Yuval Yarom}, {and} \bibinfo{person}{Mike Hamburg}.}
  \bibinfo{year}{2018}\natexlab{}.
\newblock \showarticletitle{Meltdown: Reading Kernel Memory from User Space}.
  In \bibinfo{booktitle}{\emph{27th USENIX Security Symposium (USENIX Security
  18)}}. \bibinfo{publisher}{USENIX Association}, \bibinfo{address}{Baltimore,
  MD}, \bibinfo{pages}{973--990}.
\newblock
\showISBNx{978-1-939133-04-5}
\urldef\tempurl%
\url{https://www.usenix.org/conference/usenixsecurity18/presentation/lipp}
\showURL{%
\tempurl}


\bibitem[Liu et~al\mbox{.}(2026)]%
        {liu_latticebox_2026}
\bibfield{author}{\bibinfo{person}{Zhanpeng Liu}, \bibinfo{person}{Chenyang
  Li}, \bibinfo{person}{Wende Tan}, \bibinfo{person}{Yuan Li},
  \bibinfo{person}{Xinhui Han}, \bibinfo{person}{Xi Cao}, \bibinfo{person}{Yong
  Xie}, {and} \bibinfo{person}{Chao Zhang}.} \bibinfo{year}{2026}\natexlab{}.
\newblock \showarticletitle{LatticeBox: {A} Hardware-Software Co-Designed
  Framework for Scalable and Low-Latency Compartmentalization}. In
  \bibinfo{booktitle}{\emph{33rd Annual Network and Distributed System Security
  Symposium, {NDSS} 2026, San Diego, California, USA, February 23-27, 2026}}.
  \bibinfo{publisher}{The Internet Society}.
\newblock
\urldef\tempurl%
\url{https://www.ndss-symposium.org/ndss-paper/latticebox-a-hardware-software-co-designed-framework-for-scalable-and-low-latency-compartmentalization/}
\showURL{%
\tempurl}


\bibitem[McKee et~al\mbox{.}(2022)]%
        {mckee_hakc_2022}
\bibfield{author}{\bibinfo{person}{Derrick~Paul McKee}, \bibinfo{person}{Yianni
  Giannaris}, \bibinfo{person}{Carolina Ortega}, \bibinfo{person}{Howard~E.
  Shrobe}, \bibinfo{person}{Mathias Payer}, \bibinfo{person}{Hamed Okhravi},
  {and} \bibinfo{person}{Nathan Burow}.} \bibinfo{year}{2022}\natexlab{}.
\newblock \showarticletitle{Preventing Kernel Hacks with HAKCs}. In
  \bibinfo{booktitle}{\emph{29th Annual Network and Distributed System Security
  Symposium, {NDSS} 2022, San Diego, California, USA, April 24-28, 2022}}.
  \bibinfo{publisher}{The Internet Society}.
\newblock
\urldef\tempurl%
\url{https://www.ndss-symposium.org/ndss-paper/auto-draft-257/}
\showURL{%
\tempurl}


\bibitem[Narayan et~al\mbox{.}(2020)]%
        {narayan_retrofitting_2020}
\bibfield{author}{\bibinfo{person}{Shravan Narayan}, \bibinfo{person}{Craig
  Disselkoen}, \bibinfo{person}{Tal Garfinkel}, \bibinfo{person}{Nathan Froyd},
  \bibinfo{person}{Eric Rahm}, \bibinfo{person}{Sorin Lerner},
  \bibinfo{person}{Hovav Shacham}, {and} \bibinfo{person}{Deian Stefan}.}
  \bibinfo{year}{2020}\natexlab{}.
\newblock \showarticletitle{Retrofitting Fine Grain Isolation in the Firefox
  Renderer}. In \bibinfo{booktitle}{\emph{29th USENIX Security Symposium
  (USENIX Security 20)}}. \bibinfo{publisher}{USENIX Association},
  \bibinfo{pages}{699--716}.
\newblock
\showISBNx{978-1-939133-17-5}
\urldef\tempurl%
\url{https://www.usenix.org/conference/usenixsecurity20/presentation/narayan}
\showURL{%
\tempurl}


\bibitem[Narayan et~al\mbox{.}(2023)]%
        {narayan_hfi_2023}
\bibfield{author}{\bibinfo{person}{Shravan Narayan}, \bibinfo{person}{Tal
  Garfinkel}, \bibinfo{person}{Mohammadkazem Taram}, \bibinfo{person}{Joey
  Rudek}, \bibinfo{person}{Daniel Moghimi}, \bibinfo{person}{Evan Johnson},
  \bibinfo{person}{Chris Fallin}, \bibinfo{person}{Anjo Vahldiek-Oberwagner},
  \bibinfo{person}{Michael LeMay}, \bibinfo{person}{Ravi Sahita},
  \bibinfo{person}{Dean Tullsen}, {and} \bibinfo{person}{Deian Stefan}.}
  \bibinfo{year}{2023}\natexlab{}.
\newblock \showarticletitle{Going beyond the Limits of SFI: Flexible and Secure
  Hardware-Assisted In-Process Isolation with HFI}. In
  \bibinfo{booktitle}{\emph{Proceedings of the 28th ACM International
  Conference on Architectural Support for Programming Languages and Operating
  Systems, Volume 3}} (Vancouver, BC, Canada) \emph{(\bibinfo{series}{ASPLOS
  2023})}. \bibinfo{publisher}{Association for Computing Machinery},
  \bibinfo{address}{New York, NY, USA}, \bibinfo{pages}{266–281}.
\newblock
\showISBNx{9781450399180}
\href{https://doi.org/10.1145/3582016.3582023}{doi:\nolinkurl{10.1145/3582016.3582023}}


\bibitem[Nienhuis et~al\mbox{.}(2020)]%
        {nienhuis_rigorous_2020}
\bibfield{author}{\bibinfo{person}{Kyndylan Nienhuis},
  \bibinfo{person}{Alexandre Joannou}, \bibinfo{person}{Thomas Bauereiss},
  \bibinfo{person}{Anthony Fox}, \bibinfo{person}{Michael Roe},
  \bibinfo{person}{Brian Campbell}, \bibinfo{person}{Matthew Naylor},
  \bibinfo{person}{Robert~M. Norton}, \bibinfo{person}{Simon~W. Moore},
  \bibinfo{person}{Peter~G. Neumann}, \bibinfo{person}{Ian Stark},
  \bibinfo{person}{Robert N.~M. Watson}, {and} \bibinfo{person}{Peter Sewell}.}
  \bibinfo{year}{2020}\natexlab{}.
\newblock \showarticletitle{Rigorous engineering for hardware security:
  {Formal} modelling and proof in the {CHERI} design and implementation
  process}. In \bibinfo{booktitle}{\emph{2020 {IEEE} {Symposium} on {Security}
  and {Privacy} ({SP})}}. \bibinfo{publisher}{IEEE},
  \bibinfo{pages}{1003--1020}.
\newblock
\href{https://doi.org/10.1109/SP40000.2020.00055}{doi:\nolinkurl{10.1109/SP40000.2020.00055}}


\bibitem[Roessler and DeHon(2021)]%
        {roessler_scalpel_2021}
\bibfield{author}{\bibinfo{person}{Nick Roessler} {and}
  \bibinfo{person}{Andr\'{e} DeHon}.} \bibinfo{year}{2021}\natexlab{}.
\newblock \showarticletitle{SCALPEL: Exploring the Limits of Tag-enforced
  Compartmentalization}.
\newblock \bibinfo{journal}{\emph{J. Emerg. Technol. Comput. Syst.}}
  \bibinfo{volume}{18}, \bibinfo{number}{1}, Article \bibinfo{articleno}{2}
  (\bibinfo{date}{Sept.} \bibinfo{year}{2021}), \bibinfo{numpages}{28}~pages.
\newblock
\showISSN{1550-4832}
\href{https://doi.org/10.1145/3461673}{doi:\nolinkurl{10.1145/3461673}}


\bibitem[Rugg et~al\mbox{.}(2024)]%
        {rugg_suite_2023}
\bibfield{author}{\bibinfo{person}{Peter Rugg}, \bibinfo{person}{Jonathan
  Woodruff}, \bibinfo{person}{Alexandre Joannou}, {and}
  \bibinfo{person}{Simon~W. Moore}.} \bibinfo{year}{2024}\natexlab{}.
\newblock \showarticletitle{{ A Suite of Processors to Explore CHERI-RISC-V
  Micro Architecture }}. In \bibinfo{booktitle}{\emph{2024 27th Euromicro
  Conference on Digital System Design (DSD)}}. \bibinfo{publisher}{IEEE
  Computer Society}, \bibinfo{address}{Los Alamitos, CA, USA},
  \bibinfo{pages}{351--360}.
\newblock
\href{https://doi.org/10.1109/DSD64264.2024.00054}{doi:\nolinkurl{10.1109/DSD64264.2024.00054}}


\bibitem[Schrammel et~al\mbox{.}(2022)]%
        {schrammel_jenny_2022}
\bibfield{author}{\bibinfo{person}{David Schrammel}, \bibinfo{person}{Samuel
  Weiser}, \bibinfo{person}{Richard Sadek}, {and} \bibinfo{person}{Stefan
  Mangard}.} \bibinfo{year}{2022}\natexlab{}.
\newblock \showarticletitle{Jenny: Securing Syscalls for {PKU-based} Memory
  Isolation Systems}. In \bibinfo{booktitle}{\emph{31st USENIX Security
  Symposium (USENIX Security 22)}}. \bibinfo{publisher}{USENIX Association},
  \bibinfo{address}{Boston, MA}, \bibinfo{pages}{936--952}.
\newblock
\showISBNx{978-1-939133-31-1}
\urldef\tempurl%
\url{https://www.usenix.org/conference/usenixsecurity22/presentation/schrammel}
\showURL{%
\tempurl}


\bibitem[Schrammel et~al\mbox{.}(2020)]%
        {schrammel_donky_2020}
\bibfield{author}{\bibinfo{person}{David Schrammel}, \bibinfo{person}{Samuel
  Weiser}, \bibinfo{person}{Stefan Steinegger}, \bibinfo{person}{Martin
  Schwarzl}, \bibinfo{person}{Michael Schwarz}, \bibinfo{person}{Stefan
  Mangard}, {and} \bibinfo{person}{Daniel Gruss}.}
  \bibinfo{year}{2020}\natexlab{}.
\newblock \showarticletitle{Donky: Domain Keys {\textendash} Efficient
  {In-Process} Isolation for {RISC-V} and x86}. In
  \bibinfo{booktitle}{\emph{29th USENIX Security Symposium (USENIX Security
  20)}}. \bibinfo{publisher}{USENIX Association}, \bibinfo{pages}{1677--1694}.
\newblock
\showISBNx{978-1-939133-17-5}
\urldef\tempurl%
\url{https://www.usenix.org/conference/usenixsecurity20/presentation/schrammel}
\showURL{%
\tempurl}


\bibitem[Skorstengaard et~al\mbox{.}(2019a)]%
        {skorstengaard_reasoning_2019}
\bibfield{author}{\bibinfo{person}{Lau Skorstengaard},
  \bibinfo{person}{Dominique Devriese}, {and} \bibinfo{person}{Lars Birkedal}.}
  \bibinfo{year}{2019}\natexlab{a}.
\newblock \showarticletitle{Reasoning about a Machine with Local Capabilities:
  Provably Safe Stack and Return Pointer Management}.
\newblock \bibinfo{journal}{\emph{ACM Trans. Program. Lang. Syst.}}
  \bibinfo{volume}{42}, \bibinfo{number}{1}, Article \bibinfo{articleno}{5}
  (\bibinfo{date}{Dec.} \bibinfo{year}{2019}), \bibinfo{numpages}{53}~pages.
\newblock
\showISSN{0164-0925}
\href{https://doi.org/10.1145/3363519}{doi:\nolinkurl{10.1145/3363519}}


\bibitem[Skorstengaard et~al\mbox{.}(2019b)]%
        {skorstengaard_stktokens_2019}
\bibfield{author}{\bibinfo{person}{Lau Skorstengaard},
  \bibinfo{person}{Dominique Devriese}, {and} \bibinfo{person}{Lars Birkedal}.}
  \bibinfo{year}{2019}\natexlab{b}.
\newblock \showarticletitle{StkTokens: enforcing well-bracketed control flow
  and stack encapsulation using linear capabilities}.
\newblock \bibinfo{journal}{\emph{Proc. ACM Program. Lang.}}
  \bibinfo{volume}{3}, \bibinfo{number}{POPL}, Article \bibinfo{articleno}{19}
  (\bibinfo{date}{Jan.} \bibinfo{year}{2019}), \bibinfo{numpages}{28}~pages.
\newblock
\href{https://doi.org/10.1145/3290332}{doi:\nolinkurl{10.1145/3290332}}


\bibitem[Sullivan et~al\mbox{.}(2017)]%
        {sullivan_dover_2017}
\bibfield{author}{\bibinfo{person}{Gregory~T. Sullivan},
  \bibinfo{person}{André DeHon}, \bibinfo{person}{Steven Milburn},
  \bibinfo{person}{Eli Boling}, \bibinfo{person}{Marco Ciaffi},
  \bibinfo{person}{Jothy Rosenberg}, {and} \bibinfo{person}{Andrew
  Sutherland}.} \bibinfo{year}{2017}\natexlab{}.
\newblock \showarticletitle{The Dover inherently secure processor}. In
  \bibinfo{booktitle}{\emph{2017 IEEE International Symposium on Technologies
  for Homeland Security (HST)}}. \bibinfo{pages}{1--5}.
\newblock
\href{https://doi.org/10.1109/THS.2017.7943502}{doi:\nolinkurl{10.1109/THS.2017.7943502}}


\bibitem[Tsampas et~al\mbox{.}(2019)]%
        {tsampas_temporal_2019}
\bibfield{author}{\bibinfo{person}{Stelios Tsampas}, \bibinfo{person}{Dominique
  Devriese}, {and} \bibinfo{person}{Frank Piessens}.}
  \bibinfo{year}{2019}\natexlab{}.
\newblock \showarticletitle{Temporal Safety for Stack Allocated Memory on
  Capability Machines}. In \bibinfo{booktitle}{\emph{2019 IEEE 32nd Computer
  Security Foundations Symposium (CSF)}}. \bibinfo{pages}{243--24312}.
\newblock
\showISSN{2374-8303}
\href{https://doi.org/10.1109/CSF.2019.00024}{doi:\nolinkurl{10.1109/CSF.2019.00024}}


\bibitem[Vahldiek-Oberwagner et~al\mbox{.}(2019)]%
        {vahldiek_erim_2019}
\bibfield{author}{\bibinfo{person}{Anjo Vahldiek-Oberwagner},
  \bibinfo{person}{Eslam Elnikety}, \bibinfo{person}{Nuno~O. Duarte},
  \bibinfo{person}{Michael Sammler}, \bibinfo{person}{Peter Druschel}, {and}
  \bibinfo{person}{Deepak Garg}.} \bibinfo{year}{2019}\natexlab{}.
\newblock \showarticletitle{{ERIM}: Secure, Efficient In-process Isolation with
  Protection Keys ({{{{{MPK}}}}})}. In \bibinfo{booktitle}{\emph{28th USENIX
  Security Symposium (USENIX Security 19)}}. \bibinfo{publisher}{USENIX
  Association}, \bibinfo{address}{Santa Clara, CA},
  \bibinfo{pages}{1221--1238}.
\newblock
\showISBNx{978-1-939133-06-9}
\urldef\tempurl%
\url{https://www.usenix.org/conference/usenixsecurity19/presentation/vahldiek-oberwagner}
\showURL{%
\tempurl}


\bibitem[Voulimeneas et~al\mbox{.}(2022)]%
        {voulimeneas_cerberus_2022}
\bibfield{author}{\bibinfo{person}{Alexios Voulimeneas}, \bibinfo{person}{Jonas
  Vinck}, \bibinfo{person}{Ruben Mechelinck}, {and} \bibinfo{person}{Stijn
  Volckaert}.} \bibinfo{year}{2022}\natexlab{}.
\newblock \showarticletitle{You shall not (by)pass! practical, secure, and fast
  PKU-based sandboxing}. In \bibinfo{booktitle}{\emph{Proceedings of the
  Seventeenth European Conference on Computer Systems}} (Rennes, France)
  \emph{(\bibinfo{series}{EuroSys '22})}. \bibinfo{publisher}{Association for
  Computing Machinery}, \bibinfo{address}{New York, NY, USA},
  \bibinfo{pages}{266–282}.
\newblock
\showISBNx{9781450391627}
\href{https://doi.org/10.1145/3492321.3519560}{doi:\nolinkurl{10.1145/3492321.3519560}}


\bibitem[Wahbe et~al\mbox{.}(1993)]%
        {wahbe_efficient_1993}
\bibfield{author}{\bibinfo{person}{Robert Wahbe}, \bibinfo{person}{Steven
  Lucco}, \bibinfo{person}{Thomas~E. Anderson}, {and} \bibinfo{person}{Susan~L.
  Graham}.} \bibinfo{year}{1993}\natexlab{}.
\newblock \showarticletitle{Efficient software-based fault isolation}. In
  \bibinfo{booktitle}{\emph{Proceedings of the Fourteenth ACM Symposium on
  Operating Systems Principles}} (Asheville, North Carolina, USA)
  \emph{(\bibinfo{series}{SOSP '93})}. \bibinfo{publisher}{Association for
  Computing Machinery}, \bibinfo{address}{New York, NY, USA},
  \bibinfo{pages}{203–216}.
\newblock
\showISBNx{0897916328}
\href{https://doi.org/10.1145/168619.168635}{doi:\nolinkurl{10.1145/168619.168635}}


\bibitem[Watson et~al\mbox{.}(2015)]%
        {watson_cheri_2015}
\bibfield{author}{\bibinfo{person}{Robert~N.M. Watson},
  \bibinfo{person}{Jonathan Woodruff}, \bibinfo{person}{Peter~G. Neumann},
  \bibinfo{person}{Simon~W. Moore}, \bibinfo{person}{Jonathan Anderson},
  \bibinfo{person}{David Chisnall}, \bibinfo{person}{Nirav Dave},
  \bibinfo{person}{Brooks Davis}, \bibinfo{person}{Khilan Gudka},
  \bibinfo{person}{Ben Laurie}, \bibinfo{person}{Steven~J. Murdoch},
  \bibinfo{person}{Robert Norton}, \bibinfo{person}{Michael Roe},
  \bibinfo{person}{Stacey Son}, {and} \bibinfo{person}{Munraj Vadera}.}
  \bibinfo{year}{2015}\natexlab{}.
\newblock \showarticletitle{{ CHERI: A Hybrid Capability-System Architecture
  for Scalable Software Compartmentalization }}. In
  \bibinfo{booktitle}{\emph{2015 IEEE Symposium on Security and Privacy (SP)}}.
  \bibinfo{publisher}{IEEE Computer Society}, \bibinfo{address}{Los Alamitos,
  CA, USA}, \bibinfo{pages}{20--37}.
\newblock
\showISSN{1081-6011}
\href{https://doi.org/10.1109/SP.2015.9}{doi:\nolinkurl{10.1109/SP.2015.9}}


\bibitem[Watson et~al\mbox{.}(2024)]%
        {watson_cheri_2024}
\bibfield{author}{\bibinfo{person}{Robert N.~M. Watson}, \bibinfo{person}{David
  Chisnall}, \bibinfo{person}{Jessica Clarke}, \bibinfo{person}{Brooks Davis},
  \bibinfo{person}{Nathaniel~Wesley Filardo}, \bibinfo{person}{Ben Laurie},
  \bibinfo{person}{Simon~W. Moore}, \bibinfo{person}{Peter~G. Neumann},
  \bibinfo{person}{Alexander Richardson}, \bibinfo{person}{Peter Sewell},
  \bibinfo{person}{Konrad Witaszczyk}, {and} \bibinfo{person}{Jonathan
  Woodruff}.} \bibinfo{year}{2024}\natexlab{}.
\newblock \showarticletitle{CHERI: Hardware-Enabled C/C++ Memory Protection at
  Scale}.
\newblock \bibinfo{journal}{\emph{IEEE Security and Privacy}}
  \bibinfo{volume}{22}, \bibinfo{number}{4} (\bibinfo{date}{July}
  \bibinfo{year}{2024}), \bibinfo{pages}{50–61}.
\newblock
\showISSN{1540-7993}
\href{https://doi.org/10.1109/MSEC.2024.3396701}{doi:\nolinkurl{10.1109/MSEC.2024.3396701}}


\bibitem[Watson et~al\mbox{.}(2023a)]%
        {watson_early_2023}
\bibfield{author}{\bibinfo{person}{Robert N.~M. Watson},
  \bibinfo{person}{Jessica Clarke}, \bibinfo{person}{Peter Sewell},
  \bibinfo{person}{Jonathan Woodruff}, \bibinfo{person}{Simon~W. Moore},
  \bibinfo{person}{Graeme Barnes}, \bibinfo{person}{Richard Grisenthwaite},
  \bibinfo{person}{Kathryn Stacer}, \bibinfo{person}{Silviu Baranga}, {and}
  \bibinfo{person}{Alexander Richardson}.} \bibinfo{year}{2023}\natexlab{a}.
\newblock \bibinfo{booktitle}{\emph{{Early performance results from the
  prototype Morello microarchitecture}}}.
\newblock \bibinfo{type}{{T}echnical {R}eport} UCAM-CL-TR-986.
  \bibinfo{institution}{University of Cambridge, Computer Laboratory}.
\newblock
\href{https://doi.org/10.48456/tr-986}{doi:\nolinkurl{10.48456/tr-986}}


\bibitem[Watson et~al\mbox{.}(2023b)]%
        {watson_capability_2023}
\bibfield{author}{\bibinfo{person}{Robert N.~M. Watson},
  \bibinfo{person}{Peter~G. Neumann}, \bibinfo{person}{Jonathan Woodruff},
  \bibinfo{person}{Michael Roe}, \bibinfo{person}{Hesham Almatary},
  \bibinfo{person}{Jonathan Anderson}, \bibinfo{person}{John Baldwin},
  \bibinfo{person}{Graeme Barnes}, \bibinfo{person}{David Chisnall},
  \bibinfo{person}{Jessica Clarke}, \bibinfo{person}{Brooks Davis},
  \bibinfo{person}{Lee Eisen}, \bibinfo{person}{Nathaniel~Wesley Filardo},
  \bibinfo{person}{Franz~A. Fuchs}, \bibinfo{person}{Richard Grisenthwaite},
  \bibinfo{person}{Alexandre Joannou}, \bibinfo{person}{Ben Laurie},
  \bibinfo{person}{A.~Theodore Markettos}, \bibinfo{person}{Simon~W. Moore},
  \bibinfo{person}{Steven~J. Murdoch}, \bibinfo{person}{Kyndylan Nienhuis},
  \bibinfo{person}{Robert Norton}, \bibinfo{person}{Alexander Richardson},
  \bibinfo{person}{Peter Rugg}, \bibinfo{person}{Peter Sewell},
  \bibinfo{person}{Stacey Son}, {and} \bibinfo{person}{Hongyan Xia}.}
  \bibinfo{year}{2023}\natexlab{b}.
\newblock \bibinfo{booktitle}{\emph{{Capability Hardware Enhanced RISC
  Instructions: CHERI Instruction-Set Architecture (Version 9)}}}.
\newblock \bibinfo{type}{{T}echnical {R}eport} UCAM-CL-TR-987.
  \bibinfo{institution}{University of Cambridge, Computer Laboratory}.
\newblock
\href{https://doi.org/10.48456/tr-987}{doi:\nolinkurl{10.48456/tr-987}}


\bibitem[Watson et~al\mbox{.}(2012)]%
        {watson_cheri_2012}
\bibfield{author}{\bibinfo{person}{Robert N.~M. Watson}, \bibinfo{person}{Peter
  G. Neumann~Jonathan Woodruff}, \bibinfo{person}{Jonathan Anderson},
  \bibinfo{person}{Ross Anderson}, \bibinfo{person}{Nirav Dave},
  \bibinfo{person}{Ben Laurie}, \bibinfo{person}{Simon~W. Moore},
  \bibinfo{person}{Steven~J. Murdoch}, \bibinfo{person}{Philip Paeps},
  \bibinfo{person}{Michael Roe}, {and} \bibinfo{person}{Hassen Saidi}.}
  \bibinfo{year}{2012}\natexlab{}.
\newblock \showarticletitle{{CHERI: a research platform deconflating hardware
  virtualization and protection}}. In \bibinfo{booktitle}{\emph{{Runtime
  Environments, Systems, Layering and Virtualized Environments (RESoLVE
  2012)}}}.
\newblock


\bibitem[Wesley~Filardo et~al\mbox{.}(2020)]%
        {filardo_cornucopia_2020}
\bibfield{author}{\bibinfo{person}{Nathaniel Wesley~Filardo},
  \bibinfo{person}{Brett~F. Gutstein}, \bibinfo{person}{Jonathan Woodruff},
  \bibinfo{person}{Sam Ainsworth}, \bibinfo{person}{Lucian Paul-Trifu},
  \bibinfo{person}{Brooks Davis}, \bibinfo{person}{Hongyan Xia},
  \bibinfo{person}{Edward Tomasz~Napierala}, \bibinfo{person}{Alexander
  Richardson}, \bibinfo{person}{John Baldwin}, \bibinfo{person}{David
  Chisnall}, \bibinfo{person}{Jessica Clarke}, \bibinfo{person}{Khilan Gudka},
  \bibinfo{person}{Alexandre Joannou}, \bibinfo{person}{A. Theodore~Markettos},
  \bibinfo{person}{Alfredo Mazzinghi}, \bibinfo{person}{Robert~M. Norton},
  \bibinfo{person}{Michael Roe}, \bibinfo{person}{Peter Sewell},
  \bibinfo{person}{Stacey Son}, \bibinfo{person}{Timothy~M. Jones},
  \bibinfo{person}{Simon~W. Moore}, \bibinfo{person}{Peter~G. Neumann}, {and}
  \bibinfo{person}{Robert N.~M. Watson}.} \bibinfo{year}{2020}\natexlab{}.
\newblock \showarticletitle{Cornucopia: Temporal Safety for CHERI Heaps}. In
  \bibinfo{booktitle}{\emph{2020 IEEE Symposium on Security and Privacy (SP)}}.
  \bibinfo{pages}{608--625}.
\newblock
\showISSN{2375-1207}
\href{https://doi.org/10.1109/SP40000.2020.00098}{doi:\nolinkurl{10.1109/SP40000.2020.00098}}


\bibitem[Woodruff et~al\mbox{.}(2014)]%
        {woodruff_cheri_2014}
\bibfield{author}{\bibinfo{person}{Jonathan Woodruff},
  \bibinfo{person}{Robert~N.M. Watson}, \bibinfo{person}{David Chisnall},
  \bibinfo{person}{Simon~W. Moore}, \bibinfo{person}{Jonathan Anderson},
  \bibinfo{person}{Brooks Davis}, \bibinfo{person}{Ben Laurie},
  \bibinfo{person}{Peter~G. Neumann}, \bibinfo{person}{Robert Norton}, {and}
  \bibinfo{person}{Michael Roe}.} \bibinfo{year}{2014}\natexlab{}.
\newblock \showarticletitle{The CHERI capability model: revisiting RISC in an
  age of risk}. In \bibinfo{booktitle}{\emph{Proceeding of the 41st Annual
  International Symposium on Computer Architecuture}} (Minneapolis, Minnesota,
  USA) \emph{(\bibinfo{series}{ISCA '14})}. \bibinfo{publisher}{IEEE Press},
  \bibinfo{pages}{457–468}.
\newblock
\showISBNx{9781479943944}


\bibitem[Xia et~al\mbox{.}(2019)]%
        {xia_cherivoke_2019}
\bibfield{author}{\bibinfo{person}{Hongyan Xia}, \bibinfo{person}{Jonathan
  Woodruff}, \bibinfo{person}{Sam Ainsworth}, \bibinfo{person}{Nathaniel~W.
  Filardo}, \bibinfo{person}{Michael Roe}, \bibinfo{person}{Alexander
  Richardson}, \bibinfo{person}{Peter Rugg}, \bibinfo{person}{Peter~G.
  Neumann}, \bibinfo{person}{Simon~W. Moore}, \bibinfo{person}{Robert N.~M.
  Watson}, {and} \bibinfo{person}{Timothy~M. Jones}.}
  \bibinfo{year}{2019}\natexlab{}.
\newblock \showarticletitle{CHERIvoke: Characterising Pointer Revocation using
  CHERI Capabilities for Temporal Memory Safety}. In
  \bibinfo{booktitle}{\emph{Proceedings of the 52nd Annual IEEE/ACM
  International Symposium on Microarchitecture}} (Columbus, OH, USA)
  \emph{(\bibinfo{series}{MICRO-52})}. \bibinfo{publisher}{Association for
  Computing Machinery}, \bibinfo{address}{New York, NY, USA},
  \bibinfo{pages}{545–557}.
\newblock
\showISBNx{9781450369381}
\href{https://doi.org/10.1145/3352460.3358288}{doi:\nolinkurl{10.1145/3352460.3358288}}


\bibitem[Xia et~al\mbox{.}(2018)]%
        {xia_cherirtos_2018}
\bibfield{author}{\bibinfo{person}{Hongyan Xia}, \bibinfo{person}{Jonathan
  Woodruff}, \bibinfo{person}{Hadrien Barral}, \bibinfo{person}{Lawrence
  Esswood}, \bibinfo{person}{Alexandre Joannou}, \bibinfo{person}{Robert
  Kovacsics}, \bibinfo{person}{David Chisnall}, \bibinfo{person}{Michael Roe},
  \bibinfo{person}{Brooks Davis}, \bibinfo{person}{Edward Napierala},
  \bibinfo{person}{John Baldwin}, \bibinfo{person}{Khilan Gudka},
  \bibinfo{person}{Peter~G. Neumann}, \bibinfo{person}{Alexander Richardson},
  \bibinfo{person}{Simon~W. Moore}, {and} \bibinfo{person}{Robert N.~M.
  Watson}.} \bibinfo{year}{2018}\natexlab{}.
\newblock \showarticletitle{CheriRTOS: A Capability Model for Embedded
  Devices}. In \bibinfo{booktitle}{\emph{2018 IEEE 36th International
  Conference on Computer Design (ICCD)}}. \bibinfo{pages}{92--99}.
\newblock
\showISSN{2576-6996}
\href{https://doi.org/10.1109/ICCD.2018.00023}{doi:\nolinkurl{10.1109/ICCD.2018.00023}}


\bibitem[Yang et~al\mbox{.}(2024)]%
        {yang_endokernel_2024}
\bibfield{author}{\bibinfo{person}{Fangfei Yang}, \bibinfo{person}{Bumjin Im},
  \bibinfo{person}{Weijie Huang}, \bibinfo{person}{Kelly Kaoudis},
  \bibinfo{person}{Anjo Vahldiek-Oberwagner}, \bibinfo{person}{Chia che Tsai},
  {and} \bibinfo{person}{Nathan Dautenhahn}.} \bibinfo{year}{2024}\natexlab{}.
\newblock \showarticletitle{Endokernel: A Thread Safe Monitor for Lightweight
  Subprocess Isolation}. In \bibinfo{booktitle}{\emph{33rd USENIX Security
  Symposium (USENIX Security 24)}}. \bibinfo{publisher}{USENIX Association},
  \bibinfo{address}{Philadelphia, PA}, \bibinfo{pages}{145--162}.
\newblock
\showISBNx{978-1-939133-44-1}
\urldef\tempurl%
\url{https://www.usenix.org/conference/usenixsecurity24/presentation/yang-fangfei}
\showURL{%
\tempurl}


\bibitem[Yang et~al\mbox{.}(2017)]%
        {yang_spdk_2017}
\bibfield{author}{\bibinfo{person}{Ziye Yang}, \bibinfo{person}{James~R.
  Harris}, \bibinfo{person}{Benjamin Walker}, \bibinfo{person}{Daniel Verkamp},
  \bibinfo{person}{Changpeng Liu}, \bibinfo{person}{Cunyin Chang},
  \bibinfo{person}{Gang Cao}, \bibinfo{person}{Jonathan Stern},
  \bibinfo{person}{Vishal Verma}, {and} \bibinfo{person}{Luse~E. Paul}.}
  \bibinfo{year}{2017}\natexlab{}.
\newblock \showarticletitle{SPDK: A Development Kit to Build High Performance
  Storage Applications}. In \bibinfo{booktitle}{\emph{2017 IEEE International
  Conference on Cloud Computing Technology and Science (CloudCom)}}.
  \bibinfo{pages}{154--161}.
\newblock
\showISSN{2330-2186}
\href{https://doi.org/10.1109/CloudCom.2017.14}{doi:\nolinkurl{10.1109/CloudCom.2017.14}}


\bibitem[Yedidia(2024)]%
        {yedidia_lfi_2024}
\bibfield{author}{\bibinfo{person}{Zachary Yedidia}.}
  \bibinfo{year}{2024}\natexlab{}.
\newblock \showarticletitle{Lightweight Fault Isolation: Practical, Efficient,
  and Secure Software Sandboxing}. In \bibinfo{booktitle}{\emph{Proceedings of
  the 29th ACM International Conference on Architectural Support for
  Programming Languages and Operating Systems, Volume 2}} (La Jolla, CA, USA)
  \emph{(\bibinfo{series}{ASPLOS '24})}. \bibinfo{publisher}{Association for
  Computing Machinery}, \bibinfo{address}{New York, NY, USA},
  \bibinfo{pages}{649–665}.
\newblock
\showISBNx{9798400703850}
\href{https://doi.org/10.1145/3620665.3640408}{doi:\nolinkurl{10.1145/3620665.3640408}}


\bibitem[Yee et~al\mbox{.}(2010)]%
        {yee_native_2009}
\bibfield{author}{\bibinfo{person}{Bennet Yee}, \bibinfo{person}{David Sehr},
  \bibinfo{person}{Gregory Dardyk}, \bibinfo{person}{J.~Bradley Chen},
  \bibinfo{person}{Robert Muth}, \bibinfo{person}{Tavis Ormandy},
  \bibinfo{person}{Shiki Okasaka}, \bibinfo{person}{Neha Narula}, {and}
  \bibinfo{person}{Nicholas Fullagar}.} \bibinfo{year}{2010}\natexlab{}.
\newblock \showarticletitle{Native Client: a sandbox for portable, untrusted
  x86 native code}.
\newblock \bibinfo{journal}{\emph{Commun. ACM}} \bibinfo{volume}{53},
  \bibinfo{number}{1} (\bibinfo{date}{Jan.} \bibinfo{year}{2010}),
  \bibinfo{pages}{91–99}.
\newblock
\showISSN{0001-0782}
\href{https://doi.org/10.1145/1629175.1629203}{doi:\nolinkurl{10.1145/1629175.1629203}}


\end{thebibliography}

\appendix

\section{Appendix}

The appendix is in the technical report version of this paper at \url{\TechnicalReportURL}.

\end{document}